\documentclass[12pt,a4paper]{amsart}

\usepackage[headings]{fullpage}
\usepackage{amsmath,amssymb}
\usepackage{microtype}
\UseMicrotypeSet[expansion]{basictext}
\usepackage[hidelinks]{hyperref}

\hypersetup{
  pdftitle={Purity maxima, absolute separability, and spectral volumes of absolutely PPT states},
  pdfauthor={Anh T. Tran},
  pdfsubject={Purity maxima, absolute separability of an inner polytope,
    and spectral-volume asymptotics for absolutely PPT states},
  pdfkeywords={absolutely PPT states, absolute separability, purity,
    semidefinite optimization, spectral volume}
}

\numberwithin{equation}{section}

\theoremstyle{plain}
\newtheorem{theorem}{Theorem}[section]
\newtheorem{lemma}[theorem]{Lemma}
\newtheorem{proposition}[theorem]{Proposition}
\newtheorem{corollary}[theorem]{Corollary}
\newtheorem{conjecture}[theorem]{Conjecture}
\newtheorem{maintheorem}{Theorem}

\newcommand{\BC}{\mathbb{C}}
\newcommand{\CA}{\mathcal{A}}
\newcommand{\CP}{\mathcal{P}}
\newcommand{\CR}{\mathcal{R}}
\newcommand{\CU}{\mathcal{U}}
\newcommand{\volrel}{\operatorname{vol}_{\mathrm{rel}}}
\newcommand{\CO}{\mathcal{O}}

\begin{document}

\title[Purity and spectral volumes of APPT states]%
  {Purity maxima, absolute separability, and spectral volumes of absolutely PPT states}

\author{Anh T. Tran}
\thanks{Department of Mathematical Sciences, The University of Texas at Dallas,
  Richardson, TX 75080, USA\@. Corresponding author:
  \href{mailto:att140830@utdallas.edu}{att140830@utdallas.edu}.}


\begin{abstract}
  For $m\otimes n$ systems with $n\ge m\ge3$, we determine the exact maximum
  purity of absolutely positive partial transpose (APPT) states and the
  unique maximizing ordered spectrum when $n\ge2m^2-4$ for even $m$ or
  $n\ge2m^2-m-4$ for odd $m$.  Under these dimension assumptions,
  absolutely separable states have the same maximum purity, attained at the same
  unique ordered spectrum.  The proof uses an outer polytope defined by
  two linear head--tail inequalities, each comparing a sum of the largest
  eigenvalues with a sum of the smallest.

  Under explicit multiplicity assumptions, we derive an exact matrix
  criterion for APPT membership of full-rank spectra with three distinct
  eigenvalues.
  For every $m\ge10$, this criterion gives explicit APPT boundary spectra
  at $n=4m-2$ and $n=4m$ that disprove both cases of the
  maximum purity conjecture of Ahiable, Kothakonda, and Winter.

  We also determine the exact completely absolutely separable (CAS) purity
  maximum and its unique maximizing ordered spectrum in every bipartite
  dimension.
  We prove that the inner APPT polytope consists entirely of absolutely
  separable spectra.  The resulting linear sufficient condition for absolute
  separability strictly improves the reduction-map and eigenvalue-ratio
  tests.  Finally, we compute the exact relative volumes of the CAS set and an
  outer polytope defined by a single head--tail inequality.  Under the uniform
  measure on spectra, the CAS, absolutely separable, and APPT volumes have the
  same asymptotic order
  $\sqrt n\,e^{-\kappa_m n}$ for every fixed $m\ge2$, with explicit
  $\kappa_m>0$.
\end{abstract}
\maketitle

\section{Introduction}

A bipartite quantum state is \emph{separable} if it is a convex combination
of product states; otherwise it is entangled.  A state has \emph{positive
partial transpose} (PPT) if transposing one subsystem gives a positive
semidefinite operator.  The PPT criterion is necessary for separability
\cite{Pe}.  A state $\rho$ is \emph{absolutely positive partial transpose}
(APPT) if $U\rho U^\dagger$ is PPT for every global unitary $U$, and
\emph{absolutely separable} (AS) if every such conjugate is separable.
Thus every AS state is APPT\@.  These absolute properties depend only on the
spectrum, although PPT and separability of an individual state can depend on
its eigenvectors.  Verstraete, Audenaert, and De Moor obtained the
exact criterion, in terms of the spectrum, for a $2\otimes2$ state to be
absolutely separable \cite{VAD}, while Johnston proved that APPT and AS
coincide for every $2\otimes n$ system \cite{Jo}.  Whether the
two classes coincide in all bipartite dimensions remains open \cite{AJR,AKW}.

Hildebrand characterized APPT spectra by finitely many matrix inequalities
\cite[Theorem~1]{Hi}.  In the permutation formulation of Ahiable, Kothakonda,
and Winter \cite[Theorem~4.1]{AKW}, APPT membership can be checked by testing
whether finitely many symmetric matrices, each depending linearly on the
spectrum, are positive semidefinite.  Throughout, $2\le m\le n$ are the
local dimensions and $d=mn$.

For a state $\rho$ on $\BC^m\otimes\BC^n$ with ordered spectrum
$\boldsymbol{\lambda}$, define its purity by
\[
  F(\boldsymbol{\lambda})=\operatorname{Tr}(\rho^2)
  =\sum_{i=1}^{d}\lambda_i^2.
\]
It ranges from $1/d$ at the maximally mixed state to $1$ at a pure state, and
\[
  F(\boldsymbol\lambda)=\frac1d+
  \left\|\boldsymbol\lambda-\frac1d(1,\ldots,1)\right\|_2^2.
\]
Thus maximizing purity over APPT states is equivalent to finding the APPT
spectra farthest from the maximally mixed spectrum.

The APPT purity maximum has a direct interpretation in terms of unitary
entanglement generation.  If a state's purity exceeds this maximum, then some
global unitary transforms it into a state with negative partial transpose and
hence into an entangled state.  Purity below this maximum does not, in
general, guarantee APPT or absolute separability.

For qubit--qudit systems, Song and Chen determined the common maximum
purity of absolutely separable and APPT states
\cite[Corollary~11(ii),(iv)]{SC}.  The maximum is $3/8$ for $2\otimes2$,
while for $2\otimes n$ with $n\ge3$ it equals
\[
  \begin{cases}
    \dfrac{2}{3n}, & n\text{ even},\\[6pt]
    \dfrac{6n+4}{(3n+1)^2}, & n\text{ odd}.
  \end{cases}
\]

Our upper bounds on APPT purity follow from simple scalar inequalities derived
from Hildebrand's matrix criterion.  For $2\le k\le m$, set
\[
  r_k=\frac{k(k-1)}2,
  \qquad
  s_k=\frac{k(k+1)}2.
\]
Every ordered APPT spectrum $\lambda_1\ge\cdots\ge\lambda_d$ satisfies
\begin{equation}
  \label{eq:intro-head-tail}
  \sum_{i=1}^{r_k}\lambda_i
  \le
  \sum_{j=0}^{s_k-1}\lambda_{d-j},
  \qquad 2\le k\le m.
\end{equation}
We call \eqref{eq:intro-head-tail} the \emph{$k$th head--tail inequality}:
it compares the $r_k$ largest eigenvalues, the head of the ordered spectrum,
with the $s_k$ smallest, its tail.  Write $G_k(\boldsymbol\lambda)$ for the
right-hand side of \eqref{eq:intro-head-tail} minus its left-hand side, so the
$k$th head--tail inequality is $G_k(\boldsymbol\lambda)\ge0$.
Together with ordering and normalization, selected members of this family
define polyhedral outer approximations of the ordered APPT set.  Although the
head--tail inequalities are elementary consequences of Hildebrand's criterion,
using only $G_2$ and $G_m$ together produces an outer polytope
sharp enough to determine the purity maximum in the dimension ranges of
Theorem~\ref{thm:main}.  Since
purity is strictly convex, every maximizing point of such a polytope is
a vertex.  If a maximizing vertex is APPT, the resulting upper bound
on APPT purity is attained.

We also use a known inner polytope contained in the APPT spectral set.
Jivulescu, Lupa, Nechita, and Reeb introduced this Gershgorin polytope and
proved its inclusion in APPT
\cite[Definition~7.1 and Theorem~7.2]{J+}.  Ahiable, Kothakonda, and Winter
expressed the same criterion as a permutation-invariant inner polytope,
determined its maximum purity \cite[Theorems~6.2--6.3]{AKW}, and reported
numerical comparisons with APPT optimization for $m=2,3,4$ \cite[Fig.~6]{AKW}.
Write $\Lambda^{\mathrm{APPT}}_{m,n}$ for the set of all APPT spectra on
$\BC^m\otimes\BC^n$, and let $\CO_d$ be the ordered probability simplex.
We use
\[
  \CA_{m,n}:=\Lambda^{\mathrm{APPT}}_{m,n}\cap\CO_d
\]
for the ordered APPT set and
\[
  \CP_{m,n}
  :=\left\{\boldsymbol{\lambda}\in\CO_d:
  \sum_{i=1}^{m-1}\lambda_i
  \le 2\lambda_d+\sum_{i=1}^{m-1}\lambda_{d-i}\right\}
\]
for the linear inner APPT polytope.  Thus $\CP_{m,n}\subseteq\CA_{m,n}$.
The maximum purity on $\CP_{m,n}$ is attained by one of the following two
spectra \cite[Theorem~6.3]{AKW}.  We write $x^{[q]}$ for $q$ consecutive
entries equal to $x$, with no entries when $q=0$.
The first candidate has a single larger eigenvalue:
\[
  \boldsymbol{\lambda}^{\mathrm{one}}
  =\frac1{d+2}\bigl(3,1^{[d-1]}\bigr),
  \qquad
  F(\boldsymbol{\lambda}^{\mathrm{one}})=\frac{d+8}{(d+2)^2}.
\]
For the second, put
\[
  i_\star=\left\lceil\frac{(m-1)n}{2}\right\rceil,
  \qquad D_\star=(m-1)d+2i_\star.
\]
The central spectrum is
\begin{equation}
  \label{eq:CAS-spectrum-all-dimensions}
  \boldsymbol{\lambda}^{\mathrm{cen}}_{m,n}
  =\frac1{D_\star}\bigl((m+1)^{[i_\star]},(m-1)^{[d-i_\star]}\bigr).
\end{equation}
Writing $N:=(m^2-1)n$, its purity is
\begin{equation}
  \label{eq:CAS-purity-all-dimensions}
  \begin{aligned}
    P^{\mathrm{cen}}_{m,n}
      :=F(\boldsymbol{\lambda}^{\mathrm{cen}}_{m,n})
      &=\frac{(m-1)^2d+4mi_\star}{D_\star^2}\\
      &=\begin{cases}
        \dfrac mN, & (m-1)n\text{ even},\\[6pt]
        \dfrac{m(N+2)}{(N+1)^2}, & (m-1)n\text{ odd}.
      \end{cases}
  \end{aligned}
\end{equation}
Then \cite[Theorem~6.3]{AKW} proved that
\[
  \max_{\boldsymbol{\eta}\in\CP_{m,n}}F(\boldsymbol{\eta})
  =\max\left\{\frac{d+8}{(d+2)^2},
  P^{\mathrm{cen}}_{m,n}\right\}.
\]
This leads to the following conjecture \cite[Corollary~6.4 and
Conjecture~6.7]{AKW}.

\begin{conjecture}[Ahiable, Kothakonda, and Winter]
  \label{conj:AKW}
  Let $2\le m\le n$ and $n>2$, and define
  \begin{equation*}
    n^*(m)
    :=\frac{2\bigl(m^2-2+\sqrt{(m^2-1)(m^2-4)}\bigr)}{m}.
  \end{equation*}
  Then the maximum purity over APPT spectra is
  \begin{equation*}
    \max_{\boldsymbol{\lambda}\in\Lambda^{\mathrm{APPT}}_{m,n}}
    F(\boldsymbol{\lambda})
    =
    \begin{cases}
      \dfrac{d+8}{(d+2)^2},   & n<n^*(m), \\[6pt]
      P^{\mathrm{cen}}_{m,n}, & n\ge n^*(m)
    \end{cases}.
  \end{equation*}
  The conjectured value is attained at
  $\boldsymbol{\lambda}^{\mathrm{one}}$ or
  $\boldsymbol{\lambda}^{\mathrm{cen}}_{m,n}$, respectively.
  Equivalently, the conjecture asserts that the APPT and inner-polytope
  purity maxima agree.
\end{conjecture}

We first prove that the conjectured central value $P^{\mathrm{cen}}_{m,n}$
is the exact APPT maximum when $n$ is sufficiently large relative to $m$.

\begin{maintheorem}
  \label{thm:main}
  Let $n\ge m\ge3$ be integers and suppose that
  \begin{equation}
    \label{eq:main-dimension-bounds}
    n \ge
    \begin{cases}
      2m^2-4   & \text{if $m$ is even}, \smallskip \\
      2m^2-m-4 & \text{if $m$ is odd}.
    \end{cases}
  \end{equation}
  Then the maximum purity of APPT states on
  $\BC^m\otimes\BC^n$ is $P^{\mathrm{cen}}_{m,n}$ from
  \eqref{eq:CAS-purity-all-dimensions}.  The unique maximizing ordered
  spectrum is $\boldsymbol{\lambda}^{\mathrm{cen}}_{m,n}$ from
  \eqref{eq:CAS-spectrum-all-dimensions}.
  The maximizing APPT states are
  precisely the unitary conjugates of a diagonal state with this spectrum.
\end{maintheorem}

For $m=3$, the common maximum purity of absolutely separable and APPT
states is known for every $n\ge3$ \cite[Theorem~A]{TrQ}.  The unique
maximizing ordered spectrum is $\boldsymbol\lambda^{\mathrm{one}}$ for
$3\le n\le8$ and $\boldsymbol\lambda^{\mathrm{cen}}_{3,n}$ for $n\ge9$,
where
\[
  \boldsymbol\lambda^{\mathrm{cen}}_{3,n}
  =\frac1{4n}(2^{[n]},1^{[2n]}),\qquad
  P^{\mathrm{cen}}_{3,n}=\frac3{8n}.
\]
For $m=3$, Theorem~\ref{thm:main} applies when $n\ge11$.  Thus it adds new APPT results
only for $m\ge4$, although its proof applies uniformly for $m\ge3$ without
using the qutrit purity theorem.

The dimension bounds in Theorem~\ref{thm:main} are sharp for the outer-polytope
argument, but may not be optimal for the APPT maximum itself; see
Subsection~\ref{sec:outer-sharpness}.

We next construct counterexamples to the conjectured formula in higher local
dimensions.  Under explicit multiplicity assumptions, we derive an exact
finite matrix criterion for full-rank spectra with three distinct eigenvalues
and use it to construct APPT boundary spectra with purity strictly above the
inner-polytope maximum.  When the largest eigenvalue is simple, the criterion
reduces to the two scalar inequalities in
Corollary~\ref{cor:simple-largest-appt-criterion}.

For $m\ge5$, the value $n^*(m)$ separating the two branches of the
conjectured formula satisfies
\[
  4m-2<n^*(m)<4m.
\]
Hence $n=4m-2$ and $n=4m$ fall under different branches of the
conjectured formula.  For every $m\ge10$, counterexamples occur in both
dimensions:

\begin{maintheorem}
  \label{thm:counterexample-families}
  For every integer $m\ge10$ and each $n\in\{4m-2,4m\}$,
  there is an explicit full-rank APPT boundary spectrum on
  $\BC^m\otimes\BC^n$ whose purity strictly exceeds the inner-polytope
  maximum.  Every scalar head--tail inequality is strict at this spectrum.
  Consequently, Conjecture~\ref{conj:AKW} fails in both branches.
\end{maintheorem}

Kondra et al.\ introduced \emph{completely absolutely separable} (CAS)
states, a subclass of AS states that cannot become entangled with
nonzero probability under probabilistic unital channels \cite{KBNKB}; thus
\[
  \mathrm{CAS}\subseteq\mathrm{AS}\subseteq\mathrm{APPT}.
\]
They also proved \cite[Supplemental Material, Lemma~4]{KBNKB}
that every CAS state on $\BC^m\otimes\BC^n$, with $n\ge m\ge2$, satisfies
\[
  F(\boldsymbol{\lambda})\le \frac{m}{(m^2-1)n}.
\]
Section~\ref{sec:CAS-purity} determines the exact CAS purity maximum in every
bipartite dimension.  The unique maximizing ordered spectrum is the central
spectrum \eqref{eq:CAS-spectrum-all-dimensions}.  The bound
$m/((m^2-1)n)$ is attained exactly when $(m-1)n$ is even.  Under the
hypotheses of Theorem~\ref{thm:main}, Corollary~\ref{cor:as-purity} shows that APPT, AS,
and CAS have a common maximum purity, attained at the central spectrum.

We also show that the inner APPT polytope consists entirely of absolutely
separable spectra.

\begin{maintheorem}
  \label{thm:inner-polytope-asep}
  Let $2\le m\le n$.  A state on $\BC^m\otimes\BC^n$ is absolutely
  separable if its ordered eigenvalues satisfy
  \[
    \sum_{j=1}^{m-1}\lambda_j
    \le 2\lambda_d+\sum_{j=1}^{m-1}\lambda_{d-j}.
  \]
\end{maintheorem}

Theorem~\ref{thm:inner-polytope-asep} answers the absolute-separability
question in \cite[Section~7]{AKW}.  The criterion is linear in the ordered
eigenvalues.  Proposition~\ref{prop:inner-polytope-comparison} shows that
it strictly improves the reduction-map criteria and their convex-hull
refinement \cite[Theorems~3--4]{AVMRS}, as well as the CAS ratio criterion.
For $m\ge3$, it also certifies spectra outside the convex hull of all
these earlier regions.

The inclusions $\mathrm{CAS}\subseteq\mathrm{AS}\subseteq\mathrm{APPT}$,
together with the outer bound $G_m\ge0$, also give quantitative information
about the size of these spectral sets.  Let
$\Lambda^{\mathrm{AS}}_{m,n}$ and $\Lambda^{\mathrm{CAS}}_{m,n}$ denote the
full spectral sets of AS and CAS states, respectively.  We use $\volrel$ for
Euclidean volume in the affine hyperplane of normalized spectra, normalized
by the volume of the probability simplex.

We compute the exact CAS volume and the volume of the outer polytope defined
by $G_m\ge0$.  For each fixed $m$, the ratio of the outer volume to the CAS
volume tends to a positive finite constant.  These inclusions then give the
same asymptotic order for the CAS, AS, and APPT spectral volumes.
All logarithms are natural.

\begin{maintheorem}
  \label{thm:volume-main}
  For each fixed $m\ge2$, put
  \[
    \kappa_m=m\left[
    \frac{(m+1)\log(m+1)-(m-1)\log(m-1)}2-\log2
    \right]>0.
  \]
  For $\mathsf X\in\{\mathrm{CAS},\mathrm{AS},\mathrm{APPT}\}$,
  \[
    \log\volrel(\Lambda^{\mathsf X}_{m,n})
    =-\kappa_m n+\frac12\log n+O_m(1)
    \qquad(n\to\infty).
  \]
  Equivalently,
  $\volrel(\Lambda^{\mathsf X}_{m,n})
    =\Theta_m(\sqrt n\,e^{-\kappa_m n})$.
\end{maintheorem}

We use uniform measure on spectra, which differs from the induced
probability measures on quantum states studied in \cite{CNY}.

\smallskip \noindent\textbf{Relation to earlier work.}
The well-known necessary condition for APPT
\[
  \lambda_1\le \lambda_{d-2}+\lambda_{d-1}+\lambda_d
\]
appears in \cite[Theorem~8.1(1)]{J+} and is the $k=2$
case of the head--tail family derived in Lemma~\ref{lem:head-tail}.  The earlier paper
\cite[Lemma~2.1 and Theorem~2.2]{TrS} studies the ordered polytope
$\mathcal S_d$ defined by this single inequality, classifies its vertices,
and maximizes purity on it for $d\ge8$.  The resulting bound recovers the
exact APPT maximum for every $2\otimes n$ system with $n\ge4$
\cite[Theorem~A]{TrS}.  For $3\otimes n$ systems with $n\ge3$, this
one-inequality bound equals the value conjectured by D\~ung and Kh\^oi for
the maximum APPT purity \cite[Conjecture~3.2]{DK}.  That paper shows that the
actual APPT maximum is strictly smaller \cite[Remark~1.1]{TrS}.
The exact common maximum purity of absolutely separable and APPT states
in every $3\otimes n$ system with $n\ge3$ is determined in
\cite[Theorem~A]{TrQ}.

For $m\ge3$, we add $G_m\ge0$ to the earlier condition
$G_2\ge0$.  The resulting polytope can acquire new vertices in triangular
faces of the ordered simplex, so an edge analysis alone is no longer
sufficient.  The three-eigenvalue matrix criterion also gives the boundary
spectra used to disprove Conjecture~\ref{conj:AKW}.  Wang, Chen, and Song
\cite{WCS} study only the two-qutrit case, specifically full-rank APPT
boundary points and extreme points having three distinct eigenvalues.  By
contrast, under its multiplicity assumptions, our criterion in
Lemma~\ref{lem:three-eigenvalue-appt-criterion} applies in every bipartite
dimension $n\ge m\ge3$.  For further work on boundary and extreme points of absolutely
separable and APPT states, see \cite{SC,HMS}.

For spectral volumes, Ahiable, Kothakonda, and Winter determined the exact
volume of the inner polytope \cite[Proposition~6.14]{AKW} and compared it
numerically with the APPT volume.  The relative volume of APPT spectra for
$2\otimes n$ systems is $\Theta(\sqrt n(4/27)^n)$ as $n\to\infty$,
with exponential decay rate $\kappa_2=\log(27/4)$, as proved in
\cite{TrS}.  Theorem~\ref{thm:volume-main} extends this APPT decay estimate to every fixed
$m\ge2$ and shows that the CAS, AS, and APPT spectral volumes have the same
asymptotic order.

\smallskip \noindent\textbf{Organization.}
Section~\ref{sec:preliminaries} gives the spectral and polyhedral
preliminaries, recalls Hildebrand's criterion, and derives the head--tail
inequalities.  Section~\ref{sec:proof-main} proves Theorem~\ref{thm:main}
using the two-inequality outer polytope and shows that these dimension
bounds are sharp for the outer-polytope argument.
Section~\ref{sec:counterexamples} derives, under explicit multiplicity
assumptions, an exact APPT criterion for full-rank spectra with
three distinct eigenvalues, gives a two-scalar reduction for a simple largest
eigenvalue, and constructs the counterexamples proving
Theorem~\ref{thm:counterexample-families}.  Section~\ref{sec:CAS-purity}
determines the exact CAS purity maximum in every bipartite dimension and
deduces the corresponding AS maximum under the hypotheses of
Theorem~\ref{thm:main}.  Section~\ref{sec:inner-polytope-asep}
proves Theorem~\ref{thm:inner-polytope-asep} and compares its criterion with
earlier sufficient criteria.  Section~\ref{sec:volume} gives the exact CAS
and outer-polytope volumes and proves Theorem~\ref{thm:volume-main}.

\section{Spectral preliminaries and APPT criteria}
\label{sec:preliminaries}

\subsection{Ordered spectra and polyhedral geometry}

Recall $d=mn$, and write
\[
  \mathcal H_d
  :=\left\{\boldsymbol{\lambda}\in\mathbb R^d:
  \sum_{i=1}^d\lambda_i=1\right\}
\]
for the affine hyperplane of normalized spectra.  Let
\[
  \Delta_{d-1}
  :=\left\{\boldsymbol{\lambda}\in\mathcal H_d:
  \lambda_i\ge0\right\}.
\]
Define the set of APPT spectra by
\[
  \Lambda^{\mathrm{APPT}}_{m,n}
  :=\left\{\boldsymbol{\lambda}\in\Delta_{d-1}:
  \operatorname{diag}(\boldsymbol{\lambda})\ \text{is APPT on }
  \BC^m\otimes\BC^n\right\}.
\]
The ordered simplex is
\[
  \CO_d
  :=\left\{\boldsymbol{\lambda}\in\mathcal H_d:
  \lambda_1\ge\cdots\ge\lambda_d\ge0\right\}.
\]
As in the introduction, $\CA_{m,n}=\Lambda^{\mathrm{APPT}}_{m,n}\cap\CO_d$ is
the ordered APPT set.  The sets $\Lambda^{\mathrm{AS}}_{m,n}$ and
$\Lambda^{\mathrm{CAS}}_{m,n}$ are defined in the same way using absolute
separability and complete absolute separability, respectively.

Throughout, ``APPT boundary'' means the boundary of
$\Lambda^{\mathrm{APPT}}_{m,n}$ relative to $\mathcal H_d$, rather than the
boundary of the ordered set $\CA_{m,n}$.

For $\boldsymbol\lambda\in\CO_d$, set $\lambda_{d+1}=0$ and
define the coordinates
$y_i=\lambda_i-\lambda_{i+1}$, $1\le i\le d$.  These coordinates are
nonnegative, and telescoping gives
\[
  \lambda_j=\sum_{i=j}^d y_i \quad (1\le j\le d),
  \qquad
  \sum_{i=1}^d i y_i=\sum_{j=1}^d\lambda_j=1.
\]
Conversely, any nonnegative coordinates satisfying $\sum_{i=1}^d i y_i=1$
determine a unique spectrum in $\CO_d$ by the first formula.
Thus $\boldsymbol\lambda\mapsto(y_1,2y_2,\ldots,d y_d)$ is a linear
bijection from $\CO_d$ onto $\Delta_{d-1}$.  The vertices correspond to one
coordinate $y_i=1/i$ and all other coordinates zero, giving
\[
  \boldsymbol{\nu}_i=\bigl((1/i)^{[i]},0^{[d-i]}\bigr),
  \qquad 1\le i\le d.
\]
More generally, every $\boldsymbol\lambda\in\CO_d$ has the unique
convex representation
\[
  \boldsymbol\lambda=\sum_{i=1}^d i y_i\boldsymbol\nu_i.
\]
Here $y_i>0$ exactly when $\lambda_i>\lambda_{i+1}$.
Square brackets denote the convex hull of the indicated vertices.  In
particular, $[\boldsymbol\nu_i,\boldsymbol\nu_j]$ is an edge when $i\ne j$.
A point lies in the relative interior of a simplex face exactly when its
convex combination of that face's vertices has all coefficients positive.

The \emph{rank} of a spectrum is the number of its positive entries, counted
with multiplicity.  It has full rank if all $d$ entries are positive.
The notation $x^{[q]}$ for repeated entries is as in the introduction.

Purity is strictly convex: for distinct spectra
$\boldsymbol\lambda,\boldsymbol\mu$ and $0<t<1$,
\[
  F((1-t)\boldsymbol\lambda+t\boldsymbol\mu)
  =(1-t)F(\boldsymbol\lambda)+tF(\boldsymbol\mu)
  -t(1-t)\|\boldsymbol\lambda-\boldsymbol\mu\|_2^2.
\]
In particular, every point in the relative interior of a nontrivial segment
has purity strictly below the larger endpoint purity.

Adding a linear inequality amounts to intersecting with a halfspace.
An inequality is \emph{active} at a point if it holds with equality.
We use the following face-dimension bound \cite[Lemma~2.1]{TrQ}.

\begin{lemma}
  \label{lem:active-face-dimension}
  Let $R$ be a polytope and
  $P=R\cap\{\ell_1\ge0,\ldots,\ell_s\ge0\}$, where the $\ell_\nu$ are
  linear functionals.  For $\boldsymbol\lambda\in P$, let $Q_R$ and $Q_P$
  be its smallest containing faces in $R$ and $P$, respectively.  If the
  active added functionals span a space of dimension $r$, then
  \[
    \dim Q_R\le\dim Q_P+r.
  \]
  If $R=\CO_d$ and $q=\dim Q_P$, then $\boldsymbol\lambda$ has at most
  $q+r+1$ distinct positive eigenvalues, and hence at most $r+1$ at a vertex.
  Any new vertex created by intersecting a polytope with one halfspace lies in
  the relative interior of an edge of the original polytope.
\end{lemma}

\subsection{Hildebrand's criterion and head--tail inequalities}

Recall
\[
  r_k=\frac{k(k-1)}2,\qquad s_k=\frac{k(k+1)}2,
  \qquad 2\le k\le m.
\]
We use the form of Hildebrand's criterion \cite{Hi} that allows
all permutations of the eigenvalues, as given in
\cite[Theorem~4.1]{AKW}.  Choose bijections
\[
  \begin{aligned}
    \mathfrak q&:\{(i,j):1\le i<j\le m\}
      \longrightarrow\{1,\ldots,r_m\},\\
    \mathfrak p&:\{(i,j):1\le i\le j\le m\}
      \longrightarrow\{d-s_m+1,\ldots,d\}.
  \end{aligned}
\]
The two ranges are disjoint because $r_m+s_m=m^2\le d$.
For a permutation $\pi$ of $\{1,\ldots,d\}$, let
$L_\pi(\boldsymbol{\lambda})$ be the real symmetric $m\times m$ matrix whose
upper-triangular entries are
\[
  [L_\pi(\boldsymbol{\lambda})]_{ij}=
  \begin{cases}
    2\lambda_{\pi(\mathfrak p(i,i))}, & i=j, \\
    \lambda_{\pi(\mathfrak p(i,j))}
    -\lambda_{\pi(\mathfrak q(i,j))}, & i<j.
  \end{cases}
\]
In this form, Hildebrand's criterion states that
a spectrum $\boldsymbol{\lambda}\in\Delta_{d-1}$ is APPT if and only if
\[
  L_\pi(\boldsymbol{\lambda})\succeq0\qquad\text{for every permutation }\pi.
\]
The choice of indexing bijections does not affect the result because every
permutation is tested.  We call each position $\mathfrak p(i,j)$ or
$\mathfrak q(i,j)$ a \emph{slot}.  The slots receive distinct spectral coordinates
under $\pi$, so each coordinate occurs in at most one upper-triangular
entry of $L_\pi$.

The construction reflects the fact that the partial transpose of a pure
state with Schmidt coefficients $x_1,\ldots,x_m\ge0$ has eigenvalues
\[
  x_i^2 \quad (1\le i\le m),
  \qquad \pm x_i x_j \quad (1\le i<j\le m),
\]
together with $d-m^2$ zeros \cite{AKW,Hi}.  The $\mathfrak p$-labels
correspond to the positive terms $x_i^2$ and $x_i x_j$, while the
$\mathfrak q$-labels correspond to the negative terms $-x_i x_j$.

Recall the $k$th head--tail expression
\begin{equation*}
  G_k(\boldsymbol{\lambda})
  =\sum_{i=0}^{s_k-1}\lambda_{d-i}-\sum_{i=1}^{r_k}\lambda_i.
\end{equation*}

\begin{lemma}
  \label{lem:head-tail}
  Let $\lambda_1\ge\cdots\ge\lambda_d$ be the ordered spectrum of an
  APPT state.  Then, for every integer $2\le k\le m$,
  \[
    G_k(\boldsymbol{\lambda})\ge0.
  \]
\end{lemma}

\begin{proof}
  Fix $2\le k\le m$ and let $v\in\mathbb R^m$ have its first $k$
  coordinates equal to $1$ and the rest equal to $0$.  The
  $\mathfrak p$-slots with $1\le i\le j\le k$ and the $\mathfrak q$-slots
  with $1\le i<j\le k$ use $s_k$ and $r_k$ distinct spectral coordinates,
  respectively.  Since these two sets are
  disjoint and $s_k+r_k=k^2\le d$, choose a permutation $\pi$ assigning
  the smallest $s_k$ eigenvalues to the first set and the largest $r_k$
  eigenvalues to the second; assign the remaining coordinates arbitrarily.
  Hildebrand's criterion gives $L_\pi(\boldsymbol\lambda)\succeq0$, so
  \[
    \begin{aligned}
      0\le\frac12 v^\top L_\pi(\boldsymbol\lambda)v
      &=\sum_{1\le i\le j\le k}\lambda_{\pi(\mathfrak p(i,j))}
        -\sum_{1\le i<j\le k}\lambda_{\pi(\mathfrak q(i,j))}\\
      &=\sum_{j=0}^{s_k-1}\lambda_{d-j}
        -\sum_{i=1}^{r_k}\lambda_i
       =G_k(\boldsymbol\lambda).
    \end{aligned}
  \]
  The factor $1/2$ cancels the diagonal factors of $2$ and the double
  occurrence of each off-diagonal entry in the quadratic form.
\end{proof}

Only $G_2\ge0$ and $G_m\ge0$ enter the outer polytope in
Subsection~\ref{sec:polyhedral-reduction}; the full family is used in
Section~\ref{sec:counterexamples}.

We also use the following standard sufficient condition for APPT,
proved in \cite[Theorem~7.2]{J+}.  In ordered coordinates it is the defining
inequality of the inner APPT polytope studied
in \cite[Theorem~6.2]{AKW}.

\begin{proposition}
  \label{prop:linear-appt}
  Let $\lambda_1\ge\cdots\ge\lambda_d\ge0$ be the ordered spectrum of
  a state on $\BC^m\otimes\BC^n$.  If
  \begin{equation}
    \label{eq:linear-appt}
    \sum_{i=1}^{m-1}\lambda_i
    \le 2\lambda_d+\sum_{i=1}^{m-1}\lambda_{d-i},
  \end{equation}
  then the state is APPT\@.
\end{proposition}

\section{Proof of Theorem A}
\label{sec:proof-main}

\subsection{Outer-polytope reduction}
\label{sec:polyhedral-reduction}

Except where noted, this section and Section~\ref{sec:counterexamples}
assume $3\le m\le n$.
Recall $r_m=m(m-1)/2$ and $s_m=m(m+1)/2$, and define
\[
  \CR_{m,n}:=\{\boldsymbol{\lambda}\in\CO_d:
  G_2(\boldsymbol{\lambda})\ge0,\ G_m(\boldsymbol{\lambda})\ge0\}.
\]
Proposition~\ref{prop:linear-appt} and Lemma~\ref{lem:head-tail} give
\[
  \CP_{m,n}\subseteq\CA_{m,n}\subseteq\CR_{m,n}.
\]
Thus any purity bound on $\CR_{m,n}$ is also an APPT purity bound.  Since purity is
strictly convex, every maximizing point of $\CR_{m,n}$ is a vertex.
Because only two linear inequalities are added to $\CO_d$,
Lemma~\ref{lem:active-face-dimension} shows that every vertex of
$\CR_{m,n}$ lies in a face of $\CO_d$ of dimension at most two.
A vertex in the relative interior of an edge must satisfy $G_2=0$ or
$G_m=0$, while a vertex in the relative interior of a triangular face must
satisfy both equalities.  Thus the proof reduces to four vertex classes:
the surviving simplex vertices, the $G_2$-edge intersections, the
$G_m$-edge intersections, and the triangular-face intersections.

Counting the nonzero entries of $\boldsymbol\nu_i$ in the first $r_k$
and last $s_k$ positions gives
\[
  G_k(\boldsymbol\nu_i)
    =\frac{\max\{0,i+s_k-d\}-\min\{i,r_k\}}{i}.
\]
Since $r_k+s_k=k^2\le d$ and $s_k-r_k=k$, this becomes
\begin{equation}
  \label{eq:Gk-simplex-values}
  G_k(\boldsymbol{\nu}_i)=
  \begin{cases}
    -1,        & 1\le i\le r_k, \\[2pt]
    -r_k/i,    & r_k+1\le i\le d-s_k, \\[2pt]
    (i+k-d)/i, & d-s_k+1\le i\le d.
  \end{cases}
\end{equation}
The middle range in \eqref{eq:Gk-simplex-values} is empty when $d=k^2$.
In particular, $G_k(\boldsymbol{\nu}_i)\ge0$ exactly when $i\ge d-k$.

By \eqref{eq:Gk-simplex-values}, the vertices of $\CO_d$ satisfying both
constraints are
$\boldsymbol{\nu}_{d-2},\boldsymbol{\nu}_{d-1},\boldsymbol{\nu}_{d}$. Since
$F(\boldsymbol{\nu}_i)=1/i$, their largest purity is $1/(d-2)$.

Every spectrum in $\CR_{m,n}$ has at least $d-2$ positive eigenvalues.
Indeed, if $\lambda_{d-1}=0$, then ordering gives $\lambda_d=0$, and
$G_2\ge0$ reduces to
\[
  \lambda_1\le\lambda_{d-2}.
\]
Hence $\lambda_1=\cdots=\lambda_{d-2}$, and normalization gives
$\boldsymbol\lambda=\boldsymbol\nu_{d-2}$.
Thus $\boldsymbol\nu_{d-2}$ is the only spectrum in $\CR_{m,n}$ of rank
$d-2$, and every other spectrum has rank $d-1$ or $d$.

For the $G_m$-edge maximum, we will use the assumption
\begin{equation}
  \label{eq:edge-dimension-bounds}
  n\ge
  \begin{cases}
    2m^2-4,    & m\ \text{even},\smallskip \\
    2m^2-2m-2, & m\ \text{odd}.
  \end{cases}
\end{equation}
This agrees with the hypothesis of Theorem~\ref{thm:main} when $m$ is
even and is weaker when $m$ is odd.  For odd $m$, the stronger assumption
$n\ge2m^2-m-4$ from
Theorem~\ref{thm:main} is needed only in Range~1 of
Subsection~\ref{sec:triangular-full-rank}.

For convenience, we write
\[
  p_-:=\left\lfloor\frac m2\right\rfloor,\qquad
  p_+:=\left\lceil\frac m2\right\rceil.
\]

The next two lemmas give the purity formula for spectra with two distinct
positive eigenvalues and its maximum when their ratio is
$\kappa=(m+1)/(m-1)$.  Both will also
be used in the CAS problem.

\begin{lemma}
  \label{lem:fixed-ratio-purity}
  Let $1\le i<j\le d$ be integers, let $\kappa>1$, and set
  \[
    \boldsymbol{\lambda}_{i,j,\kappa}
    =\frac{\bigl(\kappa^{[i]},1^{[j-i]},0^{[d-j]}\bigr)}
           {j+(\kappa-1)i}.
  \]
  Here $i$ is the multiplicity of the larger eigenvalue, $j$ is the rank,
  and $\kappa$ is the ratio of the two positive eigenvalues.  Then
  \[
    F(\boldsymbol{\lambda}_{i,j,\kappa})
    =:\Phi_{j,\kappa}(i)
    =\frac{j+(\kappa^2-1)i}{[j+(\kappa-1)i]^2}.
  \]
  Regard the displayed rational expression as defining
  $\Phi_{j,\kappa}(x)$ for real $0<x<j$.  For fixed $j$ and $\kappa$,
  it is strictly increasing for $x<j/(\kappa+1)$ and strictly decreasing
  for $x>j/(\kappa+1)$.  Hence
  \[
    \Phi_{j,\kappa}(x)\le
    \frac{(\kappa+1)^2}{4j\kappa},\qquad 0<x<j,
  \]
  with equality exactly when $x=j/(\kappa+1)$.
\end{lemma}

\begin{proof}
  Direct calculation gives the displayed purity formula and
  \[
    \frac{\partial\Phi_{j,\kappa}}{\partial x}
    =\frac{(\kappa-1)^2[j-(\kappa+1)x]}
           {[j+(\kappa-1)x]^3}.
  \]
  The sign of the derivative gives the monotonicity, and evaluation
  at $x=j/(\kappa+1)$ gives the maximum.
\end{proof}

For $\kappa=(m+1)/(m-1)$, we recall the maximizing
integer multiplicity from \cite[Theorem~6.3]{AKW}.  We give a direct proof
and include $m=2$ for the later CAS application.
Recall $N=(m^2-1)n$.

\begin{lemma}
  \label{lem:central-integer-maximum}
  For every $n\ge m\ge2$, put $d=mn$ and define
  \[
    P(i)=\frac{(m-1)^2d+4mi}{\bigl((m-1)d+2i\bigr)^2},
    \qquad 1\le i<d.
  \]
  The unique maximizing integer is
  $i_\star=\lceil(m-1)n/2\rceil$.  The maximum $P^{\mathrm{cen}}_{m,n}$
  equals $m/N$ when $(m-1)n$ is even and
  $m(N+2)/(N+1)^2$ when $(m-1)n$ is odd.
\end{lemma}

\begin{proof}
  Put $i_0=(m-1)n/2$.  Since $P(i)=\Phi_{d,(m+1)/(m-1)}(i)$,
  Lemma~\ref{lem:fixed-ratio-purity} shows that $P$ increases up to $i_0$
  and decreases thereafter.  The only possible maximizing integers are
  $\lfloor i_0\rfloor$ and $\lceil i_0\rceil$, both in $[1,d-1]$.
  A direct subtraction gives
  \[
    \frac mN-P(i_0+s)=\frac{4ms^2}{N(N+2s)^2}.
  \]
  If $(m-1)n$ is even, then $i_0$ is an integer and
  $P(i_0)=m/N$.  Otherwise $i_0$ is a half-integer, and the two nearest
  integers are $i_0-1/2$ and $i_0+1/2$.  In the difference formula for
  $m/N-P(i_0+s)$, the numerators are equal for $s=\pm1/2$, while the
  denominator is larger for $s=1/2$, since
  $(N+1)^2>(N-1)^2$.  Therefore
  \[
    P(i_0+1/2)>P(i_0-1/2).
  \]
  Hence $i_\star=\lceil i_0\rceil$ is the unique maximizing integer.
  Substituting $s=1/2$ gives the odd-case value.
\end{proof}

\subsection{The \texorpdfstring{$G_2$}{G2}-edge family}

\begin{proposition}
  \label{prop:G2-edges}
  Among the points in the relative interiors of edges of $\CO_d$ that
  satisfy $G_2=0$ and $G_m\ge0$, the largest value of $F$ is
  \[
    M_{G_2}=\frac{d+8p_-}{(d+2p_-)^2}.
  \]
  It is attained uniquely at the spectrum
  \[
    \frac1{d+2p_-}\bigl(3^{[p_-]},1^{[d-p_-]}\bigr).
  \]
\end{proposition}

We prove Proposition~\ref{prop:G2-edges} in two steps: first classify the
$G_2=0$ edge intersections that satisfy $G_m\ge0$, then maximize their
purity.

The signs in \eqref{eq:Gk-simplex-values} show that $G_2=0$ meets the
relative interior of $[\boldsymbol\nu_i,\boldsymbol\nu_j]$, $i<j$, exactly
when $1\le i\le d-3$ and $j\in\{d-1,d\}$.
For $1\le \ell\le d-3$, define
\[
  g_m(\ell):=-\ell G_m(\boldsymbol\nu_\ell)
  =\min\{\ell,r_m,d-m-\ell\},
\]
where the minimum representation follows from \eqref{eq:Gk-simplex-values}.

\begin{lemma}
  \label{lem:edge-ranges}
  Let $1\le i\le d-3$ and $j\in\{d-1,d\}$.
  The unique point $\boldsymbol\lambda$ in the relative interior of
  $[\boldsymbol\nu_i,\boldsymbol\nu_j]$ satisfying $G_2=0$ is
  \[
    \boldsymbol\lambda=
    \begin{cases}
      \dfrac{\bigl(2^{[i]},1^{[d-1-i]},0\bigr)}{d-1+i}, & j=d-1,\\[8pt]
      \dfrac{\bigl(3^{[i]},1^{[d-i]}\bigr)}{d+2i}, & j=d.
    \end{cases}
  \]
  Moreover, $G_m(\boldsymbol\lambda)\ge0$ if and only if
  \[
    \begin{cases}
      i\le m-1\ \text{or}\ i\ge d-2m+1, & j=d-1,\\[3pt]
      i\le p_-\ \text{or}\ i\ge d-m-p_-, & j=d.
    \end{cases}
  \]
\end{lemma}

\begin{proof}
  Write the edge spectrum as
  \[
    \boldsymbol\lambda=\bigl(a^{[i]},b^{[j-i]},0^{[d-j]}\bigr),
    \qquad a=\lambda_1>b=\lambda_{i+1}>0.
  \]
  The three smallest entries are $(b,b,0)$ when $j=d-1$ and $(b,b,b)$
  when $j=d$.  Thus $G_2=0$ gives $a=2b$ or $a=3b$, respectively.
  Normalization now gives the two displayed spectra and proves uniqueness.

  Since $\boldsymbol\lambda=i(a-b)\boldsymbol\nu_i+jb\boldsymbol\nu_j$,
  linearity and \eqref{eq:Gk-simplex-values} give
  \[
    G_m(\boldsymbol\lambda)=(j-d+m)b-g_m(i)(a-b)
    =\begin{cases}
      \dfrac{m-1-g_m(i)}{d-1+i}, & j=d-1,\\[8pt]
      \dfrac{m-2g_m(i)}{d+2i}, & j=d.
    \end{cases}
  \]
  Hence $G_m\ge0$ is equivalent to $g_m(i)\le m-1$ or $g_m(i)\le m/2$,
  respectively.  Since $g_m(i)=\min\{i,r_m,d-m-i\}$ and
  $r_m>m-1\ge m/2$, these conditions give exactly the index ranges in the lemma;
  in the second case, we use that $i$ is an integer.
\end{proof}

\begin{proof}[Proof of Proposition~\ref{prop:G2-edges}]
  \emph{Maximization at fixed rank.}
  For the spectrum in Lemma~\ref{lem:edge-ranges},
  Lemma~\ref{lem:fixed-ratio-purity} gives
  \[
    F(\boldsymbol\lambda)=\Phi_{j,\kappa}(i),
    \qquad
    \kappa=
    \begin{cases}
      2, & j=d-1,\\
      3, & j=d.
    \end{cases}
  \]
  For each rank $j$, Lemma~\ref{lem:edge-ranges} gives exactly the
  integer indices in the two intervals $1\le i\le i_L$ and
  $i_R\le i\le d-3$, where
  \[
    (i_L,i_R)=
    \begin{cases}
      (m-1,d-2m+1), & j=d-1, \\
      (p_-,d-m-p_-), & j=d.
    \end{cases}
  \]
  In both cases $i_L<j/(\kappa+1)$ and $i_R\ge j/2>j/(\kappa+1)$.
  The inequality for $i_L$ follows from $d=mn$ and $n\ge m\ge3$.
  For $i_R$, we have
  \[
    2i_R-j=
    \begin{cases}
      m(n-m)+(m-1)(m-3)\ge0,                  & j=d-1, \\
      d-2m-2p_-\ge m(n-3)\ge0, & j=d.
    \end{cases}
  \]
  By Lemma~\ref{lem:fixed-ratio-purity}, the purity increases on the first
  interval and decreases on the second.  Its only possible maxima are
  therefore at $i_L$ and $i_R$.

  Since $i_R>i_L$, clearing the positive denominators gives
  \[
    \Phi_{j,\kappa}(i_L)>\Phi_{j,\kappa}(i_R)
    \quad\Longleftrightarrow\quad
    (\kappa^2-1)i_Li_R>j(j-i_L-i_R).
  \]
  Since $i_L+i_R=d-m$ and $i_R\ge j/2$, it suffices to show
  $(\kappa^2-1)i_L>2(j-d+m)$.  Indeed,
  \[
    (\kappa^2-1)i_L-2(j-d+m)=
    \begin{cases}
      m-1>0, & j=d-1,\\
      8p_--2m\ge2m-4>0, & j=d.
    \end{cases}
  \]
  Thus, for each rank $j$, the maximum is attained uniquely at $i=i_L$.

  \smallskip \noindent\emph{Comparison of the two ranks.}
  The two maximum values are
  \[
    \Phi_{d-1,2}(m-1)=\frac{d+3m-4}{(d+m-2)^2},
    \qquad
    \Phi_{d,3}(p_-)=\frac{d+8p_-}{(d+2p_-)^2}.
  \]
  Since $(m-1)/2\le p_-<d/4$, monotonicity from
  Lemma~\ref{lem:fixed-ratio-purity} gives
  \[
    \Phi_{d,3}(p_-)
    \ge \Phi_{d,3}\!\left(\frac{m-1}{2}\right)
    =\frac{d+4m-4}{(d+m-1)^2}.
  \]
  Put $x=d+m-2\ge4m-2$.  Subtracting gives
  \[
    \begin{aligned}
      &\Phi_{d,3}(p_-)-\Phi_{d-1,2}(m-1)\\
      &\quad\ge
      \frac{x\bigl[(m-2)x-(4m-3)\bigr]-2(m-1)}
           {x^2(x+1)^2}
      \ge\frac{d-m}{x^2(x+1)^2}>0.
    \end{aligned}
  \]
  Here the second inequality uses $(m-2)x-(4m-3)\ge1$, which follows
  from $m\ge3$ and $x\ge4m-2$.
  Consequently, the unique maximizing indices are $(i,j)=(p_-,d)$,
  which give the spectrum and purity in the proposition.
\end{proof}

\subsection{The \texorpdfstring{$G_m$}{Gm}-edge family}
\label{sec:Gm-edge}

\begin{proposition}
  \label{prop:Gm-edges}
  Assume \eqref{eq:edge-dimension-bounds}.  Among the points in the relative
  interiors of edges of $\CO_d$ that satisfy $G_m=0$ and $G_2\ge0$, the
  largest value of $F$ is $P^{\mathrm{cen}}_{m,n}$.
  It is attained uniquely at the central spectrum
  $\boldsymbol\lambda^{\mathrm{cen}}_{m,n}$ in
  \eqref{eq:CAS-spectrum-all-dimensions}.
\end{proposition}

We first bound the multiplicity of the largest eigenvalue at a
$G_m=0$ edge intersection.

\begin{lemma}
  \label{lem:Gm-edge-multiplicity}
  Let $n\ge m\ge3$, and let $\boldsymbol\lambda$ lie in the relative
  interior of an edge $[\boldsymbol\nu_i,\boldsymbol\nu_j]$ of $\CO_d$,
  with $i<j$.  If $G_m(\boldsymbol\lambda)=0$ and
  $G_2(\boldsymbol\lambda)\ge0$, then
  \[
    j\in\{d-1,d\},\qquad i\ge p_+.
  \]
  In particular, the largest eigenvalue has multiplicity at least $p_+$.
\end{lemma}

\begin{proof}
  The spectrum belongs to $\CR_{m,n}$ and is not a simplex vertex.
  The rank observation in Subsection~\ref{sec:polyhedral-reduction} therefore
  gives $j\in\{d-1,d\}$.  The signs in \eqref{eq:Gk-simplex-values}
  and $G_m=0$ then imply $i<d-m<j$.

  Write $\boldsymbol\lambda=(a^{[i]},b^{[j-i]},0^{[d-j]})$ with
  $a>b>0$.  As in the proof of Lemma~\ref{lem:edge-ranges},
  \[
    G_m=0\quad\Longrightarrow\quad
    g_m(i)(a-b)=(j-d+m)b.
  \]
  For $j=d-1$, the condition $G_2\ge0$ gives $a\le2b$, and hence
  $g_m(i)\ge m-1$.  For $j=d$, it gives $a\le3b$, and hence
  $g_m(i)\ge m/2$.  Since $g_m(i)\le i$, both cases imply
  $i\ge m/2$, or equivalently $i\ge p_+$.
\end{proof}

We now maximize over the larger set of spectra with $G_m\ge0$ and a
largest eigenvalue of multiplicity at least $p_+$.  The resulting purity
bound will also apply to triangular-face vertices whose largest eigenvalue
has multiplicity at least $p_+$.

\begin{lemma}
  \label{lem:repeated-head-purity}
  Assume $n\ge m\ge3$ and \eqref{eq:edge-dimension-bounds}.  Then
  \[
    \max_{\substack{\boldsymbol\lambda\in\CO_d,\ G_m(\boldsymbol\lambda)\ge0\\
      \lambda_1=\cdots=\lambda_{p_+}}}F(\boldsymbol\lambda)
      =P^{\mathrm{cen}}_{m,n},
  \]
  with equality only at $\boldsymbol\lambda^{\mathrm{cen}}_{m,n}$.
\end{lemma}

We prove Lemma~\ref{lem:repeated-head-purity} by reducing the lower-rank
vertices to full rank and then treating the three multiplicity ranges.
Throughout this subsection, assume \eqref{eq:edge-dimension-bounds} unless
noted otherwise.

\subsubsection{Rank reduction}
\label{sec:Gm-rank-reduction}

The condition $\lambda_1=\cdots=\lambda_{p_+}$ is equivalent to
$y_1=\cdots=y_{p_+-1}=0$.  Hence these ordered spectra form the
face $[\boldsymbol\nu_{p_+},\ldots,\boldsymbol\nu_d]$ of $\CO_d$.
By \eqref{eq:Gk-simplex-values}, its vertices satisfying $G_m\ge0$ are
$\boldsymbol\nu_{d-m},\ldots,\boldsymbol\nu_d$, whose purities are
at most $1/(d-m)$.  Every other vertex after imposing $G_m\ge0$ is the unique
$G_m=0$ intersection on an edge $[\boldsymbol\nu_i,\boldsymbol\nu_j]$
with
\[
  p_+\le i<d-m<j\le d,
\]
by Lemma~\ref{lem:active-face-dimension}.
For fixed $i$, we write the intersection at rank $j$ as
\[
  \boldsymbol\lambda=t_j\boldsymbol\nu_i+(1-t_j)\boldsymbol\nu_j.
\]
Since $g_m(i)>0$, the equation $G_m=0$ becomes
\[
  -\frac{t_jg_m(i)}i+\frac{(1-t_j)(j-d+m)}j=0,
  \qquad
  t_j=\frac{i(j-d+m)}{j\,g_m(i)+i(j-d+m)}.
\]
Using $G_m=0$, we obtain
\[
  F(\boldsymbol\lambda)=\frac{t_j^2}{i}+\frac{1-t_j^2}{j},
  \qquad
  \frac{1-t_j}{j}=\frac{1-(1+g_m(i)/i)t_j}{d-m}.
\]
Consequently, $F(\boldsymbol\lambda)=Q_i(t_j)$, where
\begin{equation}
  \label{eq:rank-parameter-purity}
  Q_i(t):=\frac1{d-m}-\frac{g_m(i)}{i(d-m)}t
    +\frac{d-m-i-g_m(i)}{i(d-m)}t^2.
\end{equation}
The quadratic $Q_i$ is convex because $g_m(i)\le d-m-i$.
Moreover,
\[
  \frac{t_j}{1-t_j}
  =\frac{i}{g_m(i)}\left(1-\frac{d-m}{j}\right)
\]
increases with $j$.  Thus
\begin{equation}
  \label{eq:rank-parameter-interval}
  0<t_j\le t_*(i):=\frac{mi}{d\,g_m(i)+mi},
\end{equation}
with equality only at $j=d$.  Substituting $t=0$ into $Q_i(t)$ gives
$Q_i(0)=F(\boldsymbol\nu_{d-m})=1/(d-m)$; substituting $t=t_*(i)$ gives
the full-rank intersection on $[\boldsymbol\nu_i,\boldsymbol\nu_d]$.
For $0\le t\le t_*(i)$, convexity gives
\begin{equation}
  \label{eq:rank-convexity-bound}
  Q_i(t)\le\left(1-\frac{t}{t_*(i)}\right)Q_i(0)
    +\frac{t}{t_*(i)}Q_i(t_*(i)).
\end{equation}
It therefore suffices to prove
\[
  Q_i(0)<P^{\mathrm{cen}}_{m,n},\qquad
  Q_i(t_*(i))\le P^{\mathrm{cen}}_{m,n}.
\]
Once these bounds hold, the positive coefficient of $Q_i(0)$ makes the
inequality strict for every $t<t_*(i)$, in particular for every rank $j<d$.
This bound for real $t$ will also be used in the triangular-face analysis.

\subsubsection{Full-rank bounds and the maximum}

To bound $Q_i(t_*(i))$, substitute $t_*(i)$ from
\eqref{eq:rank-parameter-interval} into
$\boldsymbol\lambda=t_*(i)\boldsymbol\nu_i+(1-t_*(i))\boldsymbol\nu_d$:
\[
  \boldsymbol\lambda
  =\frac{\bigl((g_m(i)+m)^{[i]},g_m(i)^{[d-i]}\bigr)}
         {d\,g_m(i)+mi}.
\]
The formula for $g_m(i)$ gives three ranges.  We first analyze the central
range, which contains the maximizer, and then bound the two outer ranges.

\smallskip \noindent\textbf{Case 1: $r_m<i\le d-s_m$.}
Here $g_m(i)=r_m$, so
\[
  \boldsymbol\lambda
  =\frac{\bigl((m+1)^{[i]},(m-1)^{[d-i]}\bigr)}{(m-1)d+2i}.
\]
Its purity is $\Phi_{d,(m+1)/(m-1)}(i)$.  By
Lemma~\ref{lem:central-integer-maximum}, this is at most
$P^{\mathrm{cen}}_{m,n}$, with equality only at $i=i_\star$.
This index lies in the central range: assumption
\eqref{eq:edge-dimension-bounds} gives $n\ge m+2$, and hence
\[
  \frac{(m-1)n}{2}-r_m\ge2,
  \qquad
  d-s_m-\frac{(m-1)n}{2}\ge4,
\]
so $r_m<i_\star\le d-s_m$.  The central spectrum therefore belongs to
the full-rank family just considered and satisfies $G_m=0$.
If $y$ is its smaller positive entry, then its smallest three entries
all equal $y$ and
\[
  G_2(\boldsymbol\lambda^{\mathrm{cen}}_{m,n})
  =\left(3-\frac{m+1}{m-1}\right)y>0.
\]
Thus $P^{\mathrm{cen}}_{m,n}$ is attained by the central $G_m=0$ edge vertex.

\smallskip \noindent\textbf{Case 2: $p_+\le i\le r_m$.}
Here $g_m(i)=i$, so the spectrum has $i$ entries equal to
$(1+m/i)/(d+m)$ and $d-i$ equal to $1/(d+m)$.
Its purity decreases with $i$, giving
\[
  F(\boldsymbol\lambda)=\frac{d+2m+m^2/i}{(d+m)^2}
  \le B_{G_m}:=\frac{d+2m+m^2/p_+}{(d+m)^2}.
\]
Here $B_{G_m}$ is the purity of the full-rank $G_m=0$
intersection on $[\boldsymbol\nu_{p_+},\boldsymbol\nu_d]$.

\smallskip \noindent\textbf{Case 3: $d-s_m<i<d-m$.}
Here $g_m(i)=d-m-i$, so the quadratic term of $Q_i$ in
\eqref{eq:rank-parameter-purity} vanishes.  At the full-rank intersection,
\[
  F(\boldsymbol\lambda)=Q_i(t_*(i))
  =\frac1{d-m}-\frac{g_m(i)}{i(d-m)}t_*(i)
  <\frac1{d-m}.
\]

It remains to show that the bounds in Cases~2 and~3 are strictly below
$P^{\mathrm{cen}}_{m,n}$.  The following estimates will also be used in the
triangular-face analysis.

\begin{lemma}
  \label{lem:auxiliary-bounds}
  Assume $n\ge m\ge3$ and \eqref{eq:edge-dimension-bounds}.  Then
  \[
    \frac1{d-m}<B_{G_m}<P^{\mathrm{cen}}_{m,n},
    \qquad M_{G_2}\le B_{G_m}.
  \]
  The inequality $M_{G_2}\le B_{G_m}$ is an equality exactly when $m$ is even.
\end{lemma}

\begin{proof}
  Lemma~\ref{lem:central-integer-maximum} gives
  $P^{\mathrm{cen}}_{m,n}\ge(m/N)(1-1/(N+1)^2)$ in both parity cases.
  To compare $B_{G_m}$ with this lower bound, observe
  \[
    \frac mN\left(1-\frac1{(n+1)^2}\right)-B_{G_m}
    =\begin{cases}
      \dfrac{n-(2m^2-4)}{m(m^2-1)(n+1)^2}, & m\text{ even},\\[8pt]
      \dfrac{n-(2m^2-2m-2)}{m(m^2-1)(n+1)^2}, & m\text{ odd}.
    \end{cases}
  \]
  The right-hand side is nonnegative by \eqref{eq:edge-dimension-bounds}.
  Since $N>n$, it follows that
  \[
    B_{G_m}
    \le\frac mN\left(1-\frac1{(n+1)^2}\right)
    <\frac mN\left(1-\frac1{(N+1)^2}\right)
    \le P^{\mathrm{cen}}_{m,n}.
  \]
  For the lower comparison, $p_+\le2m/3$ and
  \eqref{eq:edge-dimension-bounds} implies $n\ge10$.  Hence
  \[
    B_{G_m}\ge\frac{d+7m/2}{(d+m)^2}
      =\frac1{d-m}+\frac{m(d-9m)}{2(d-m)(d+m)^2}
      >\frac1{d-m}.
  \]
  For the $G_2$-edge comparison, $p_+=p_-=m/2$ gives
  $B_{G_m}=M_{G_2}$ when $m$ is even.
  When $m$ is odd, \eqref{eq:edge-dimension-bounds} implies $n\ge3m$, and
  \[
    B_{G_m}-M_{G_2}
    =\frac{2mn(n-3m)+(3m+1)(n-2m+2)}
      {m(m+1)(n+1)^2(d+m-1)^2}>0.\qedhere
  \]
\end{proof}

\begin{proof}[Proof of Lemma~\ref{lem:repeated-head-purity}]
  By Subsection~\ref{sec:Gm-rank-reduction}, every vertex of the set in
  the lemma is either a simplex vertex
  $\boldsymbol\nu_{d-m},\ldots,\boldsymbol\nu_d$ or a $G_m=0$ edge
  intersection with $p_+\le i<d-m<j\le d$.
  The simplex vertices have purity at most
  $1/(d-m)<P^{\mathrm{cen}}_{m,n}$ by Lemma~\ref{lem:auxiliary-bounds}.

  Cases~1--3 give
  $F(\boldsymbol\lambda)\le P^{\mathrm{cen}}_{m,n}$ when $j=d$.
  The bound is attained only in the central range, at $i=i_\star$,
  by Lemma~\ref{lem:central-integer-maximum}.
  The rank reduction in Subsection~\ref{sec:Gm-rank-reduction}
  then gives $F(\boldsymbol\lambda)<P^{\mathrm{cen}}_{m,n}$
  for every edge vertex of rank $j<d$.

  The central spectrum satisfies $G_m=0$ and $i_\star>r_m\ge p_+$,
  so it belongs to the class considered in the lemma and attains the bound.
  It is therefore the only maximizing vertex.  Strict convexity of
  purity excludes every nonvertex, proving uniqueness.
\end{proof}

\begin{proof}[Proof of Proposition~\ref{prop:Gm-edges}]
  Lemma~\ref{lem:Gm-edge-multiplicity} places every $G_m=0$ edge
  intersection satisfying $G_2\ge0$ in the class considered in
  Lemma~\ref{lem:repeated-head-purity}.  Hence its purity is at most
  $P^{\mathrm{cen}}_{m,n}$, with equality only at the central spectrum.
  The Case~1 analysis above shows that the central spectrum is itself a
  $G_m=0$ edge point and satisfies $G_2>0$.  Hence it is the unique maximizer
  among these edge vertices.
\end{proof}

\subsection{Triangular-face intersections}

\begin{proposition}
  \label{prop:two-faces}
  Assume \eqref{eq:main-dimension-bounds}.  Every vertex of $\CR_{m,n}$ lying in
  the relative interior of a two-dimensional face of $\CO_d$ has purity strictly
  smaller than $P^{\mathrm{cen}}_{m,n}$.
\end{proposition}

Let $i$ denote the multiplicity of the largest eigenvalue at a
triangular-face vertex.  The case $i\ge p_+$ follows directly from
Lemma~\ref{lem:repeated-head-purity}.  We therefore focus on $i<p_+$.
For these vertices, convexity first reduces the rank-$(d-1)$ case to the
full-rank case.  We then split the full-rank analysis into the three ranges
of the piecewise formula for $g_m(j)$, where $j$ counts the entries larger
than the smallest positive eigenvalue.  Assumption
\eqref{eq:edge-dimension-bounds} suffices except in Range~1 for odd $m$.

\subsubsection{Geometry and preliminary reductions}

Assume \eqref{eq:edge-dimension-bounds}, and let $\boldsymbol\lambda$ be a
vertex of $\CR_{m,n}$ lying in the relative interior of a triangular face
$[\boldsymbol\nu_i,\boldsymbol\nu_j,\boldsymbol\nu_k]$ of $\CO_d$,
where $i<j<k$.  By Lemma~\ref{lem:active-face-dimension}, both added
inequalities are active:
$G_2(\boldsymbol\lambda)=G_m(\boldsymbol\lambda)=0$.
Write
\[
  \boldsymbol\lambda
  =\bigl(a^{[i]},b^{[j-i]},c^{[k-j]},0^{[d-k]}\bigr),
  \qquad a=\lambda_1>b=\lambda_{i+1}>c=\lambda_{j+1}>0.
\]
Its rank is $k$, and its simplex representation is
\[
  \boldsymbol\lambda=i(a-b)\boldsymbol\nu_i
    +j(b-c)\boldsymbol\nu_j+kc\boldsymbol\nu_k.
\]

The rank observation in Subsection~\ref{sec:polyhedral-reduction} gives
$k\in\{d-1,d\}$, since $\boldsymbol\lambda$ has three distinct positive
eigenvalues and therefore differs from $\boldsymbol\nu_{d-2}$.

\smallskip \noindent\textbf{The case $i\ge p_+$.}
The first $p_+$ entries agree, and $G_m=0$, so
Lemma~\ref{lem:repeated-head-purity} gives
$F(\boldsymbol\lambda)\le P^{\mathrm{cen}}_{m,n}$.
Equality in that lemma occurs only at the central spectrum,
whereas $\boldsymbol\lambda$ has three distinct positive values.  Therefore
\[
  F(\boldsymbol\lambda)<P^{\mathrm{cen}}_{m,n}.
\]

\smallskip \noindent\textbf{The case $i<p_+$.}
We may now assume $i<p_+$, equivalently $i<m/2$.
Then \eqref{eq:Gk-simplex-values} gives
\[
  G_m(\boldsymbol\nu_i)=-1,\qquad
  G_2(\boldsymbol\nu_i)=-\frac1i.
\]
To prove $j\le d-3$, consider the linear combination
\[
  J:=G_m-iG_2,
\]
chosen so that $J(\boldsymbol\nu_i)=0$.  At the three simplex vertices
$\boldsymbol\nu_{d-2}$, $\boldsymbol\nu_{d-1}$, and $\boldsymbol\nu_d$,
\[
  J(\boldsymbol\nu_{d-2})=\frac{m-2}{d-2}>0,\qquad
  J(\boldsymbol\nu_{d-1})=\frac{m-1-i}{d-1}>0,\qquad
  J(\boldsymbol\nu_d)=\frac{m-2i}{d}>0.
\]
Since $J(\boldsymbol\lambda)=J(\boldsymbol\nu_i)=0$, the simplex
representation gives
$j(b-c)J(\boldsymbol\nu_j)+kcJ(\boldsymbol\nu_k)=0$.
Both coefficients are positive, and $J(\boldsymbol\nu_k)>0$ because
$k\in\{d-1,d\}$.  Thus $J(\boldsymbol\nu_j)<0$, which implies
\[
  j\le d-3.
\]

Put $\zeta=d-k\in\{0,1\}$, so $\zeta=1$ means rank $d-1$ and
$\zeta=0$ means full rank.  Since $j\le d-3$, the three smallest
entries are $(c,c,0)$ when $\zeta=1$ and $(c,c,c)$ when $\zeta=0$.
Therefore $G_2=0$ gives
\[
  a=(3-\zeta)c.
\]
Let $F_\zeta(i,j)$ denote the purity of this vertex.  Thus
$F_1(i,j)$ is the rank-$(d-1)$ value and $F_0(i,j)$ is the full-rank
value.

Evaluating $G_m$ in this representation using
\eqref{eq:Gk-simplex-values}, the equation $G_m=0$ becomes
\begin{equation}
  \label{eq:triangular-Gm-relation}
  i(a-b)+g_m(j)(b-c)=(m-\zeta)c.
\end{equation}
Normalization gives
\begin{equation}
  \label{eq:triangular-normalization}
  ia+(j-i)b+(d-j-\zeta)c=1.
\end{equation}
Together with $a=(3-\zeta)c$, these equations determine $a,b,c$ once
$i,j,\zeta$ are fixed.

\subsubsection{Rank-$(d-1)$ reduction}
Suppose $\zeta=1$, so $a=2c$.  Equation~\eqref{eq:triangular-Gm-relation}
then gives
\[
  i\frac{a-b}{c}+g_m(j)\frac{b-c}{c}=m-1.
\]
Here $(a-b)/c$ and $(b-c)/c$ are positive and sum to $1$. Since $i<m/2<m-1$,
it follows that $g_m(j)>m-1$, and hence
\[
  m\le j\le d-2m.
\]

\smallskip \noindent\emph{The comparison segment.}
Fix these indices $i,j$.  Substituting $\zeta c=3c-a$ into
\eqref{eq:triangular-Gm-relation} and
\eqref{eq:triangular-normalization} gives the affine line
\begin{equation}
  \label{eq:triangular-comparison-line}
  \begin{aligned}
    (i-1)(a-c)+(g_m(j)-i)(b-c)&=(m-2)c,\\
    (i+1)a+(j-i)b+(d-j-3)c&=1.
  \end{aligned}
\end{equation}
The two equations are independent: the first has a positive coefficient
of $b$ and a negative coefficient of $c$, whereas all three coefficients
in the second are positive.

Intersect this line with $0\le c\le b\le a\le3c$.  The normalization
equation makes the intersection bounded, and the original rank-$(d-1)$
point satisfies all the ordering inequalities strictly.  Thus the
intersection is a nontrivial compact segment.  It cannot meet $c=0$, since
that would force $a=b=c=0$.  It cannot meet $b=c$ either: the first equation
would give
\[
  (m-2)c=(i-1)(a-c)\le2(i-1)c<(m-2)c.
\]
Its two endpoints therefore satisfy $a=b$ and $a=3c$, respectively.
The endpoint $a=3c$ is the full-rank point with the same $i,j$, while the
original point $a=2c$ lies in the open segment between them.

\smallskip \noindent\emph{A convex comparison.}
Using $\zeta c=3c-a$, the purity at the original rank-$(d-1)$ point
($\zeta=1$) and at the full-rank endpoint ($\zeta=0$) can both be written
as the auxiliary quadratic
\[
  \mathcal F(a,b,c):=ia^2+(j-i)b^2+(d-j-3)c^2+ac.
\]
Thus $\mathcal F=F_1(i,j)$ at the original point and
$\mathcal F=F_0(i,j)$ at the full-rank endpoint; on the rest of the segment,
$\mathcal F$ is used only as an auxiliary quadratic.  Completing the square,
\[
  \mathcal F(a,b,c)
  =i\left(a+\frac{c}{2i}\right)^2+(j-i)b^2
    +\left(d-j-3-\frac1{4i}\right)c^2.
\]
All three coefficients are positive, because $j>i\ge1$ and
$d-j-3-1/(4i)\ge2m-3-1/4>0$.  Hence $\mathcal F$ is strictly convex.

\smallskip \noindent\emph{The endpoint $a=b$.}
Set $t:=j(a-c)>0$ at this endpoint.
Equations~\eqref{eq:triangular-comparison-line}
give
\[
  (g_m(j)-1)t=j(m-2)c,
  \qquad
  (d-m)c+\left(1+\frac{g_m(j)}j\right)t=1.
\]
Since $g_m(j)>m-1$, the first equality gives $c>g_m(j)t/(mj)$.
The second then implies
\[
  0<t<\frac{mj}{d\,g_m(j)+mj}=t_*(j).
\]
Also, using
\[
  (d-m)c+\left(1+\frac{g_m(j)}j\right)t=1
\]
to eliminate $c$ gives
\[
  \mathcal F(a,a,c)=\frac{t^2}{j}+(1+t)c=Q_j(t).
\]
Since $p_+\le j<d-m$, the convexity bound
\eqref{eq:rank-convexity-bound} and its endpoint estimates give
$\mathcal F(a,a,c)<P^{\mathrm{cen}}_{m,n}$.

Strict convexity on the comparison segment now implies
\begin{equation}
  \label{eq:triangular-rank-reduction}
  F_1(i,j)<\max\{F_0(i,j),P^{\mathrm{cen}}_{m,n}\}.
\end{equation}
Thus it remains only to prove $F_0(i,j)<P^{\mathrm{cen}}_{m,n}$.

\subsubsection{Full-rank vertices}
\label{sec:triangular-full-rank}
For full rank, $\zeta=0$ and $a=3c$, so
\[
  \boldsymbol\lambda=\bigl((3c)^{[i]},b^{[j-i]},c^{[d-j]}\bigr),
  \qquad 1\le i<m/2,\qquad c<b<3c.
\]
The formula for $g_m(j)$ changes at $r_m$ and $d-s_m$, so the
full-rank argument naturally divides into
\[
  j\le r_m,\qquad r_m<j\le d-s_m,\qquad d-s_m<j\le d-3.
\]

\smallskip \noindent\textbf{Range 1: $j\le r_m$.}
Here every entry after the first $r_m$ equals $c$.
Rather than optimize separately over the triangular vertices in this range,
we enlarge the class under consideration.  Every such vertex satisfies
$G_m=0$, $G_2\ge0$, and
$\lambda_{r_m+1}=\cdots=\lambda_d=c$, so it suffices to bound the purity
of all ordered spectra satisfying $G_m=0$, $G_2\ge0$, and
$\lambda_{r_m+1}=\cdots=\lambda_d$.  The equality $G_m=0$
and normalization give
\[
  \sum_{\ell=1}^{r_m}\lambda_\ell=s_m c,
  \qquad 1=(d+m)c.
\]
Thus $c=1/(d+m)>0$.  Set
$e_\ell=\lambda_\ell/c-1$ for $1\le\ell\le r_m$.  Since the last three
entries equal $c$, the inequality $G_2\ge0$ gives $\lambda_1\le3c$.
Together with ordering and $G_m=0$, this gives
\[
  0\le e_\ell\le2,\qquad \sum_{\ell=1}^{r_m}e_\ell=m,
  \qquad
  F(\boldsymbol\lambda)
  =c^2\left(d+2m+\sum_{\ell=1}^{r_m}e_\ell^2\right).
\]
Because both the constraint set and the purity are invariant under
permutations of the $e_\ell$, ordering is irrelevant for this maximization.
Strict convexity forces every maximizer to be a vertex of the resulting
polytope.  At such a vertex, at most one coordinate can lie in $(0,2)$:
otherwise two interior coordinates could be perturbed in opposite
directions while preserving all constraints.

The condition $\sum e_\ell=m$ therefore forces $m/2$ entries to equal $2$
when $m$ is even.  When $m$ is odd, it forces $(m-1)/2$ entries to equal
$2$, one entry to equal $1$, and the rest to equal $0$.  Hence the unique
ordered maximizer of the enlarged problem is
\[
  \frac1{d+m}
  \begin{cases}
    \bigl(3^{[m/2]},1^{[d-m/2]}\bigr), & m\text{ even},\\[4pt]
    \bigl(3^{[(m-1)/2]},2,1^{[d-(m+1)/2]}\bigr), & m\text{ odd}.
  \end{cases}
\]
For odd $m$, this is the triangular-face spectrum with
$(i,j)=((m-1)/2,(m+1)/2)$; for even $m$, it has two distinct eigenvalues.
Therefore
\[
  F(\boldsymbol\lambda)\le
  \begin{cases}
    \dfrac{d+4m}{(d+m)^2}, & m\text{ even},\\[6pt]
    \dfrac{d+4m-1}{(d+m)^2}, & m\text{ odd}.
  \end{cases}
\]
For even $m$, $p_+=m/2$, so the bound equals
$B_{G_m}<P^{\mathrm{cen}}_{m,n}$ by Lemma~\ref{lem:auxiliary-bounds}.
For odd $m$, the Range~1 bound is larger than $B_{G_m}$.
Comparing it directly with $P^{\mathrm{cen}}_{m,n}$ is the only step
requiring the stronger assumption in Theorem~\ref{thm:main}.  Write
\[
  n=2m^2-m-4+\omega,\qquad \omega\ge0.
\]
Since $m$ is odd, $(m-1)n$ is even; hence
Lemma~\ref{lem:central-integer-maximum} gives
$P^{\mathrm{cen}}_{m,n}=m/[(m^2-1)n]$.  Direct subtraction gives
\[
  P^{\mathrm{cen}}_{m,n}-\frac{d+4m-1}{(d+m)^2}
  =\frac{(mn-1)\omega+m(m-1)^2+4}
  {m^2(m^2-1)n(n+1)^2}>0.
\]
Thus every full-rank triangular vertex in Range~1 has purity below
$P^{\mathrm{cen}}_{m,n}$.

For the remaining ranges, multiply normalization by $b+c$ and subtract
the purity to obtain
\begin{equation}
  \label{eq:triangular-purity-identity}
  F_0(i,j)=b+c-(d+2i)bc+6ic^2.
\end{equation}

\smallskip \noindent\textbf{Range 2: $r_m<j\le d-s_m$.}
Here $g_m(j)=r_m$.  Substituting $\zeta=0$ and $a=3c$ into
\eqref{eq:triangular-Gm-relation} gives
\[
  3ic+(r_m-i)b=s_m c.
\]
Thus, for fixed $i$, the ratio $b/c$ is independent of $j$.  Set
\[
  x_i:=\frac{b-c}{b+c}=\frac{m-2i}{m^2-4i},
  \qquad \sigma_{i,j}:=b+c.
\]
For fixed $m,n$, $x_i$ depends only on $i$, while $\sigma_{i,j}$ also depends
on $j$.  Since $1\le i<m/2$, we have $0<x_i<1/m$.  Moreover,
\[
  b=\frac{1+x_i}{2}\sigma_{i,j},\qquad
  c=\frac{1-x_i}{2}\sigma_{i,j},\qquad
  i=\frac{m(1-mx_i)}{2(1-2x_i)}.
\]
Substitution into \eqref{eq:triangular-purity-identity} gives
\begin{equation}
  \label{eq:range2-sigma-purity}
  F_0(i,j)=\sigma_{i,j}
    -\frac{N+E_{m,n}(x_i)}{4m}\sigma_{i,j}^2,
\end{equation}
where $N=(m^2-1)n$ and
\[
  E_{m,n}(x):=(1-mx)\bigl[n(1+mx)-2m^2(1-x)\bigr].
\]
Since $E_{m,n}(x_i)$ is affine in $n$ with coefficient
$1-m^2x_i^2>0$, it increases with $n$.  The hypothesis
\eqref{eq:edge-dimension-bounds} implies $n\ge2m^2-2m-2$;
evaluating at $n=2m^2-2m-2$ gives
\[
  E_{m,n}(x_i)\ge
  \frac{4i(m+1)(m-2)^2\bigl[m^2-2(m+1)i\bigr]}{(m^2-4i)^2}.
\]
Since $i$ is an integer with $1\le i<m/2$, we have
$m^2-2(m+1)i\ge1$ and $0<m^2-4i\le m^2-4$.
Consequently,
\[
  E_{m,n}(x_i)\ge\frac{4(m+1)}{(m+2)^2}
    >\frac1{m+2}>\frac1{N+2}.
\]
Maximizing the quadratic in \eqref{eq:range2-sigma-purity} over
$\sigma_{i,j}>0$ therefore gives
\[
  F_0(i,j)\le\frac{m}{N+E_{m,n}(x_i)}
    <\frac{m}{N+1/(N+2)}
    =\frac{m(N+2)}{(N+1)^2}.
\]
Lemma~\ref{lem:central-integer-maximum} gives
$\frac{m(N+2)}{(N+1)^2}\le P^{\mathrm{cen}}_{m,n}$, and hence
$F_0(i,j)<P^{\mathrm{cen}}_{m,n}$.

\smallskip \noindent\textbf{Range 3: $d-s_m<j\le d-3$.}
Here $g_m(j)=d-m-j$.  Equation~\eqref{eq:triangular-Gm-relation} and
normalization give
\[
  (g_m(j)-i)(b-c)=(m-2i)c,
  \qquad
  (d-m-2i)b+6ic=1.
\]
Since $b>c$ and $m-2i>0$, the first equation implies $g_m(j)>i$.
Also $j>d-s_m$ gives $g_m(j)<r_m$, so
\[
  \frac bc
  =1+\frac{m-2i}{g_m(j)-i}
  >1+\frac{m-2i}{r_m-i}
  =\frac{s_m-3i}{r_m-i}.
\]
The last expression is the Range~2 value of $b/c$ at $j=d-s_m$.
Together with $b<3c$, this places $b/c$ strictly between that ratio and $3$.
For fixed $i$, the second equation gives
\[
  c=\frac1{(d-m-2i)(b/c)+6i},
\]
which decreases with $b/c$ because $d-m-2i>0$.
Writing $c_{i,j}$ for the smallest eigenvalue at indices $i,j$, we obtain
\[
  \frac1{3(d-m)}<c_{i,j}<c_{i,d-s_m}.
\]
At the upper endpoint $c=c_{i,d-s_m}$,
$\Psi_i(c)=F_0(i,d-s_m)<P^{\mathrm{cen}}_{m,n}$ by Range~2.

Eliminating $b$ from the normalization equation in
\eqref{eq:triangular-purity-identity}, define for fixed $i$
\[
  \Psi_i(c)
  :=\frac{1-(m+10i)c+6i(2d-m)c^2}{d-m-2i}.
\]
Then
\[
  F_0(i,j)=\Psi_i(c_{i,j}).
\]
Since $d-m-2i>0$, $\Psi_i$ is a strictly convex quadratic, and
\[
  \Psi_i\!\left(\frac1{3(d-m)}\right)
  =\frac1{d-m}-\frac{m}{3(d-m)^2}
  <\frac1{d-m}<P^{\mathrm{cen}}_{m,n}
\]
by Lemma~\ref{lem:auxiliary-bounds}.  Since both endpoint values of
$\Psi_i$ are below $P^{\mathrm{cen}}_{m,n}$, convexity gives
\[
  F_0(i,j)<P^{\mathrm{cen}}_{m,n}.
\]

\begin{proof}[Proof of Proposition~\ref{prop:two-faces}]
  Under \eqref{eq:main-dimension-bounds},
  Lemma~\ref{lem:repeated-head-purity} and its equality condition give
  $F(\boldsymbol\lambda)<P^{\mathrm{cen}}_{m,n}$ when $i\ge p_+$.
  For $i<p_+$, the three full-rank ranges give
  $F_0(i,j)<P^{\mathrm{cen}}_{m,n}$, and
  \eqref{eq:triangular-rank-reduction} then gives
  $F_1(i,j)<P^{\mathrm{cen}}_{m,n}$.  These cases cover all the triangular-face vertices.
\end{proof}

\begin{proof}[Proof of Theorem~\ref{thm:main}]
  We compare the four vertex classes identified in
  Subsection~\ref{sec:polyhedral-reduction}.  The simplex vertices satisfy
  \[
    F(\boldsymbol\lambda)\le\frac1{d-2}
    <\frac1{d-m}<P^{\mathrm{cen}}_{m,n},
  \]
  by \eqref{eq:Gk-simplex-values} and
  Lemma~\ref{lem:auxiliary-bounds}.  The $G_2$-edge vertices satisfy
  \[
    F(\boldsymbol\lambda)\le M_{G_2}<P^{\mathrm{cen}}_{m,n}
  \]
  by Proposition~\ref{prop:G2-edges} and
  Lemma~\ref{lem:auxiliary-bounds}.
  Proposition~\ref{prop:Gm-edges} gives a unique $G_m$-edge vertex of
  purity $P^{\mathrm{cen}}_{m,n}$, while
  Proposition~\ref{prop:two-faces} places every triangular-face vertex
  strictly below it.  Hence the central spectrum is the only maximizing
  vertex of $\CR_{m,n}$.  Strict convexity excludes every nonvertex, so
  it is the unique maximizer on $\CR_{m,n}$.

  For the central spectrum, $(m-1)\lambda_1=(m+1)\lambda_d$, so
  \[
    \sum_{i=1}^{m-1}\lambda_i
    \le(m-1)\lambda_1=(m+1)\lambda_d
    \le2\lambda_d+\sum_{i=1}^{m-1}\lambda_{d-i}.
  \]
  Thus it belongs to $\CP_{m,n}\subseteq\CA_{m,n}$.
  Since $\CA_{m,n}\subseteq\CR_{m,n}$, the central spectrum is also the
  unique maximizer on $\CA_{m,n}$.  Unitary invariance gives all
  maximizing states.
\end{proof}

\subsection{Sharpness for the outer polytope}
\label{sec:outer-sharpness}

The preceding proof determines the maximum on the entire outer polytope
$\CR_{m,n}$.  We now show that the dimension bounds in
\eqref{eq:main-dimension-bounds} are sharp for that outer-polytope statement.
Let $\epsilon_m=0$ for even $m$ and $\epsilon_m=1$ for odd $m$, and take
\[
  n=2m^2-5-\epsilon_m m,
\]
one less than the corresponding bound.
The Range~1 maximizers in Subsection~\ref{sec:triangular-full-rank}
satisfy $G_2=G_m=0$ for every $n\ge m$, independently of the bound used to
compare their purities.
For this value of $n$, their purity is
$(d+4m-\epsilon_m)/(d+m)^2$.
Since $n>m^2$ and Lemma~\ref{lem:central-integer-maximum} gives
$P^{\mathrm{cen}}_{m,n}\le m/[(m^2-1)n]$, direct subtraction gives
\[
  \begin{aligned}
    \frac{d+4m-\epsilon_m}{(d+m)^2}-P^{\mathrm{cen}}_{m,n}
    &\ge\frac{d+4m-\epsilon_m}{(d+m)^2}-\frac{m}{(m^2-1)n}\\
    &=\frac{(m+\epsilon_m)n-m^3}
      {m^2(m^2-1)n(n+1)^2}>0.
  \end{aligned}
\]
Thus the central spectrum does not maximize purity on $\CR_{m,n}$ one
integer below the corresponding bound in \eqref{eq:main-dimension-bounds}.
The APPT maximum may nevertheless equal the central purity for smaller values
of $n$.

\section{Counterexamples to the AKW conjecture}
\label{sec:counterexamples}

We first derive, under explicit multiplicity assumptions, an exact finite
matrix criterion for full-rank APPT spectra with three distinct eigenvalues.
For a simple largest eigenvalue, it reduces to two scalar inequalities.
The boundary equality gives the spectra used to prove
Theorem~\ref{thm:counterexample-families}.

Recall that $r_m$ and $s_m$ are, respectively, the numbers of
$\mathfrak q$-slots and $\mathfrak p$-slots in an $m\times m$
Hildebrand matrix.

\subsection{APPT criterion for three distinct eigenvalues}

\begin{lemma}
  \label{lem:three-eigenvalue-appt-criterion}
  Let $n\ge m\ge3$, put $d=mn$, and let
  \[
    \boldsymbol\lambda
      =\frac1S\bigl(A^{[\alpha]},B^{[\beta]},C^{[\gamma]}\bigr),
    \qquad A>B>C>0,
  \]
  where $\alpha,\beta,\gamma$ are positive integers with
  $\alpha+\beta+\gamma=d$,
  \[
    S=\alpha A+\beta B+\gamma C,
  \]
  and
  \[
    \gamma\ge s_m,
    \qquad
    \alpha+\beta\ge r_m.
  \]
  Put $a_*:=\min\{\alpha,r_m\}$.  For every subset
  $E\subseteq\{(i,j):1\le i<j\le m\}$ with $|E|=a_*$, let
  $H_E\in\mathbb R^{m\times m}$ be the symmetric matrix with
  \[
    (H_E)_{ii}=2C
  \]
  and, for $i<j$,
  \[
    (H_E)_{ij}=(H_E)_{ji}
      =\begin{cases}
        C-A,& (i,j)\in E,\\
        C-B,& (i,j)\notin E.
      \end{cases}
  \]
  Then
  \[
    \boldsymbol\lambda\text{ is APPT}
    \quad\Longleftrightarrow\quad
    H_E\succeq0\quad\text{for every such }E.
  \]
  When these conditions hold, $\boldsymbol\lambda$ lies on the APPT
  boundary if and only if at least one $H_E$ is singular.
\end{lemma}

\begin{proof}
  \smallskip \noindent\emph{Necessity.}
  Fix $E\subseteq\{(i,j):1\le i<j\le m\}$ with $|E|=a_*$.  Assign $C$ to all
  $\mathfrak p$-slots, $A$ to the $\mathfrak q$-slots indexed by $E$,
  and $B$ to the remaining $\mathfrak q$-slots.  The assumptions
  $\gamma\ge s_m$ and $\alpha+\beta\ge r_m$ provide enough copies:
  the number of $B$-slots is $r_m-a_*\le\beta$ (if $\alpha<r_m$, this
  follows from $\alpha+\beta\ge r_m$, while if $\alpha\ge r_m$, the
  left-hand side is zero).
  Complete the assignment arbitrarily to a permutation $\pi_E$.  Then
  \begin{equation}
    \label{eq:matrix-realization}
    S L_{\pi_E}(\boldsymbol\lambda)=H_E.
  \end{equation}
  Thus Hildebrand's criterion implies $H_E\succeq0$ for every such $E$.

  \smallskip \noindent\emph{Sufficiency.}
  Fix a permutation $\pi$ and put $\widetilde L=S L_\pi(\boldsymbol\lambda)$.
  Every scaled spectral entry is at least $C$, so
  \[
    \widetilde L_{ii}=2S\lambda_{\pi(\mathfrak p(i,i))}\ge2C.
  \]
  Let $E_A$ consist of the off-diagonal pairs whose two defining slots
  contain at least one copy of $A$.  Distinct pairs use disjoint spectral
  coordinates, so $|E_A|\le\min\{\alpha,r_m\}=a_*$.  Enlarge $E_A$ to a
  set $E\subseteq\{(i,j):1\le i<j\le m\}$ with $|E|=a_*$.  Since an off-diagonal entry
  is the difference of its two slot values,
  \[
    |\widetilde L_{ij}|\le
    \begin{cases}
      A-C,&(i,j)\in E,\\
      B-C,&(i,j)\notin E.
    \end{cases}
  \]
  The second bound holds because neither slot then contains $A$.
  For $x\in\mathbb R^m$, put $|x|=(|x_1|,\ldots,|x_m|)^\top$.
  Using $\widetilde L_{ij}x_i x_j\ge-|\widetilde L_{ij}|\,|x_i||x_j|$ gives
  \begin{equation}
    \label{eq:quadratic-comparison}
    \begin{aligned}
      x^\top\widetilde L x
      &\ge2C\sum_{i=1}^m|x_i|^2
        -2(A-C)\sum_{(i,j)\in E}|x_i||x_j|\\
      &\qquad-2(B-C)\sum_{(i,j)\notin E}|x_i||x_j|\\
      &=|x|^\top H_E |x|.
    \end{aligned}
  \end{equation}
  If every $H_E\succeq0$, this is nonnegative for every $x$ and $\pi$,
  proving that $\boldsymbol\lambda$ is APPT\@.

  \smallskip \noindent\emph{Boundary.}
  Suppose some $H_E$ is singular and choose $0\ne v\in\ker H_E$.
  By \eqref{eq:matrix-realization}, the linear functional
  \[
    \ell(\boldsymbol\xi):=v^\top L_{\pi_E}(\boldsymbol\xi)v
  \]
  is nonnegative on all APPT spectra and vanishes at $\boldsymbol\lambda$.
  At the maximally mixed spectrum $\boldsymbol\nu_d$, it has value
  \[
    \ell(\boldsymbol\nu_d)=\frac2d\|v\|^2>0.
  \]
  Hence $\ell$ is nonconstant on $\mathcal H_d$.  Since
  $\ell(\boldsymbol\lambda)=0$, every neighborhood of
  $\boldsymbol\lambda$ in $\mathcal H_d$ contains points where $\ell<0$.
  Because $\boldsymbol\lambda$ has full rank, the neighborhood may be chosen
  inside $\Delta_{d-1}$.  Such points lie outside the APPT set, so
  $\boldsymbol\lambda$ is a boundary point.

  Conversely, if every $H_E$ is positive definite, then
  \eqref{eq:quadratic-comparison} gives
  \[
    x^\top S L_\pi(\boldsymbol\lambda)x
      \ge |x|^\top H_E |x|>0\qquad(x\ne0).
  \]
  Therefore every $L_\pi(\boldsymbol\lambda)$ is positive definite.
  All entries of $\boldsymbol\lambda$ are positive, the matrices depend
  continuously on the spectrum, and only finitely many permutations occur.
  By continuity, these strict inequalities hold throughout a sufficiently
  small neighborhood of $\boldsymbol\lambda$ in $\mathcal H_d$ that remains
  in $\Delta_{d-1}$; this neighborhood lies inside the APPT set.
  Thus $\boldsymbol\lambda$ is not on its boundary.
\end{proof}

\subsection{Boundary spectra with a simple largest eigenvalue}

When the largest eigenvalue is simple, $a_*=1$, so the general criterion
involves only the matrices $H_E$ associated with single pairs
$E=\{(p,q)\}$.  These matrices differ only by simultaneous permutations of
their rows and columns, so they have the same eigenvalues.
We now reduce their positive semidefiniteness to two scalar inequalities.

\begin{corollary}
  \label{cor:simple-largest-appt-criterion}
  Let $n\ge m\ge3$, put $d=mn$, and let
  \[
    \boldsymbol\lambda=\frac1S(A,B^{[h-1]},C^{[d-h]}),\qquad
    A>B>C>0,
  \]
  where $h$ is an integer with $r_m\le h\le d-s_m$, and
  \[
    S=A+(h-1)B+(d-h)C.
  \]
  Put
  \[
    \delta:=B-C,
    \qquad
    \chi:=2C-(m-3)\delta.
  \]
  Then
  \[
    \boldsymbol\lambda\text{ is APPT}
    \quad\Longleftrightarrow\quad
    \chi>0,
    \qquad
    (3C-A)\chi\ge2(m-2)\delta^2.
  \]
  When these conditions hold, $\boldsymbol\lambda$ lies on the APPT
  boundary if and only if
  \[
    (3C-A)\chi=2(m-2)\delta^2.
  \]
\end{corollary}

\begin{proof}
  Apply Lemma~\ref{lem:three-eigenvalue-appt-criterion} with
  $\alpha=1$, $\beta=h-1$, and $\gamma=d-h$, and fix $E=\{(p,q)\}$.

  \smallskip \noindent\emph{Positive eigenspaces.}
  Let $e_1,\ldots,e_m$ be the standard basis of $\mathbb R^m$, and set
  \[
    W_E:=\left\{x\in\mathbb R^m:x_p=x_q=0,
      \ \sum_{j\notin\{p,q\}}x_j=0\right\}.
  \]
  Direct multiplication gives
  \[
    H_E(e_p-e_q)=(A+C)(e_p-e_q),\qquad
    H_Ex=(B+C)x\quad(x\in W_E).
  \]
  Since $\dim W_E=m-3$ (with $W_E=\{0\}$ when $m=3$), these positive
  eigenspaces have total dimension $m-2$.

  \smallskip \noindent\emph{The remaining two-dimensional subspace.}
  Their orthogonal complement consists of vectors constant on $\{p,q\}$
  and on its complement.  It has the orthonormal basis
  \[
    f_E=\frac{e_p+e_q}{\sqrt2},\qquad
    g_E=\frac{\sum_{j\notin\{p,q\}}e_j}{\sqrt{m-2}}.
  \]
  In the basis $(f_E,g_E)$, the action of $H_E$ on this subspace
  is represented by
  \[
    \begin{pmatrix}
      3C-A&-\sqrt{2(m-2)}\,\delta\\
      -\sqrt{2(m-2)}\,\delta&\chi
    \end{pmatrix}.
  \]
  Since $\delta>0$, positive semidefiniteness requires $\chi>0$:
  a zero diagonal entry would force the corresponding off-diagonal entry
  to vanish.  With $\chi>0$, the determinant condition is exactly
  $(3C-A)\chi\ge2(m-2)\delta^2$, which also implies $3C-A>0$.
  This proves the two scalar conditions.

  The other invariant subspaces are positive definite, so $H_E$ is
  singular exactly when this determinant is zero.
  Lemma~\ref{lem:three-eigenvalue-appt-criterion} now gives both
  the APPT criterion and the boundary characterization.
\end{proof}

Once the multiplicity bounds hold, the APPT conditions depend only on the
ratios $A:B:C$, not on $h$.  Thus purity can be optimized over $h$ without
changing the APPT test.

To parametrize the boundary equality, set
\[
  t:=\frac{\chi}{2\delta}>0.
\]
A common positive rescaling of $A,B,C$ changes neither $t$ nor the
normalized spectrum.  We may therefore take $\delta=B-C=1$, so
$\chi=2t$.  The equality in
Corollary~\ref{cor:simple-largest-appt-criterion} then gives
\begin{equation}
  \label{eq:boundary-eigenvalues}
  C_{m,t}=\frac{m+2t-3}{2},\qquad
  B_{m,t}=C_{m,t}+1,\qquad
  A_{m,t}=3C_{m,t}-\frac{m-2}{t}.
\end{equation}
Moreover,
\[
  A_{m,t}-B_{m,t}
  =\frac{(t-1)(m+2t-2)}{t},
\]
so $A>B>C$ is equivalent to $t>1$.

The endpoint $t=1$ gives $A=B=(m+1)/2$ and $C=(m-1)/2$, so the boundary family
meets the two-eigenvalue family in
Subsection~\ref{sec:Gm-edge}.  For $t>1$,
the largest eigenvalue is simple, while for fixed $h$ the normalized spectrum
tends to $\boldsymbol\lambda^{\mathrm{one}}$ as $t\to\infty$.

\begin{proposition}
  \label{prop:boundary-spectra}
  Let $n\ge m\ge3$, let $d=mn$, and let $t>1$.  If the integer $h$
  satisfies
  \[
    r_m\le h\le d-s_m,
  \]
  define
  \[
    \boldsymbol{\lambda}_{m,n,h,t}
    =\frac1{S_{m,n,h,t}}
    \bigl(A_{m,t},B_{m,t}^{[h-1]},C_{m,t}^{[d-h]}\bigr),
  \]
  where
  \[
    S_{m,n,h,t}
    =A_{m,t}+(h-1)B_{m,t}+(d-h)C_{m,t}.
  \]
  Then $\boldsymbol{\lambda}_{m,n,h,t}$ is a full-rank APPT boundary
  spectrum outside $\CP_{m,n}$.  Moreover,
  every scalar head--tail inequality is strict, and
  \[
    \min_{2\le k\le m}
    S_{m,n,h,t}G_k(\boldsymbol\lambda_{m,n,h,t})
      =\frac{m-2}{t}\min\{1,(t-1)^2\}>0.
  \]
\end{proposition}

\begin{proof}
  \smallskip \noindent\emph{APPT boundary property.}
  Write $\boldsymbol\lambda=\boldsymbol\lambda_{m,n,h,t}$,
  $S=S_{m,n,h,t}$, and $A=A_{m,t}$, $B=B_{m,t}$, $C=C_{m,t}$.
  By \eqref{eq:boundary-eigenvalues}, these values satisfy
  $A>B>C>0$, $\delta=B-C=1$, $\chi=2t>0$, and
  $(3C-A)\chi=2(m-2)$.  Corollary~\ref{cor:simple-largest-appt-criterion}
  therefore shows that $\boldsymbol\lambda$ is a full-rank APPT boundary
  spectrum.

  \smallskip \noindent\emph{Strict head--tail inequalities.}
  For every $2\le k\le m$, we have $r_k\le r_m\le h$ and
  $s_k\le s_m\le d-h$.  In the scaled spectrum $S\boldsymbol\lambda$,
  the $r_k$ largest entries are $A$ and $r_k-1$ copies of $B$, while
  the $s_k$ smallest entries all equal $C$.  Substituting into $G_k$ gives
  \[
    \begin{aligned}
      SG_k(\boldsymbol\lambda)
        &=\frac{m-k+(k-2)(t-1)^2}{t}
          +\frac{(k-2)(m-k)}2\\
        &\ge\frac{m-2}{t}\min\{1,(t-1)^2\}>0.
    \end{aligned}
  \]
  The inequality uses $m-k,k-2\ge0$ and $(m-k)+(k-2)=m-2$.
  The lower bound is attained at $k=2$ when $t\ge2$ and at $k=m$
  when $1<t\le2$, which proves the minimum formula.

  \smallskip \noindent\emph{Outside the inner polytope.}
  Since $h\ge r_m\ge m-1$, the $m-1$ largest entries are $A/S$ and
  $m-2$ copies of $B/S$.  Since $d-h\ge s_m\ge m$, the smallest
  entries occurring on the right-hand side of \eqref{eq:linear-appt}
  are all equal to $C/S$.  The left-hand side minus the right-hand side
  of that inequality, multiplied by $S$, is
  \[
    A+(m-2)B-(m+1)C
      =\frac{(m-2)(t-1)}{t}>0.
  \]
  Hence $\boldsymbol\lambda\notin\CP_{m,n}$.
\end{proof}

\subsection{Explicit counterexample families}

We now choose parameters that give counterexamples to both cases of
Conjecture~\ref{conj:AKW}.  Recall that $n^*(m)$ is the larger
root of
\[
  q_m(x)=mx^2-4(m^2-2)x+4m.
\]
For $m\ge5$,
\[
  q_m(4m-2)=-8(m^2-5m+2)<0,
  \qquad
  q_m(4m)=36m>0.
\]
Since $q_m$ has positive leading coefficient, these signs imply
\[
  4m-2<n^*(m)<4m.
\]
Hence $n=4m-2$ and $n=4m$ fall under different branches of the
conjectured formula.

For the family used in Theorem~\ref{thm:counterexample-families}, we choose
$t=7/2$.  After a common rescaling, this choice gives integer values for
the three entries and, with the multiplicity chosen below, works for both
$n=4m-2$ and $n=4m$.  Multiplying the three values in
\eqref{eq:boundary-eigenvalues} by $14$ gives
\[
  (A_m,B_m,C_m)
  =(17m+92,\,7m+42,\,7m+28).
\]
Set
\[
  h_m:=2m^2-4m.
\]
For $m\ge10$ and $n\in\{4m-2,4m\}$, define
\[
  S_{m,n}
  :=A_m+(h_m-1)B_m+(mn-h_m)C_m
  =(7m+28)mn+28m^2-46m+50
\]
and
\[
  \boldsymbol\lambda_{m,n}
  :=\frac1{S_{m,n}}
  \bigl(A_m,B_m^{[h_m-1]},C_m^{[mn-h_m]}\bigr).
\]
\begin{proof}[Proof of Theorem~\ref{thm:counterexample-families}]
  Put $d=mn$.

  \smallskip \noindent\emph{Multiplicity conditions.}
  The chosen multiplicity satisfies the hypotheses of
  Proposition~\ref{prop:boundary-spectra}, since
  \[
    h_m-r_m=\frac{m(3m-7)}2>0,
  \]
  and
  \[
    d-h_m-s_m
      =\frac{m(2n-5m+7)}2
      \ge\frac{3m(m+1)}2>0
  \]
  for $n\in\{4m-2,4m\}$.  Proposition~\ref{prop:boundary-spectra}
  therefore gives full rank, the APPT boundary property, and strictness
  of every scalar head--tail inequality.

  \smallskip \noindent\emph{Purity comparison.}
  By the inner-polytope maximum formula in the introduction and
  $4m-2<n^*(m)<4m$, the two inner-polytope maxima are
  \[
    \begin{cases}
      \dfrac{4m^2-2m+8}{(4m^2-2m+2)^2}, & n=4m-2,\\[6pt]
      \dfrac1{4(m^2-1)}, & n=4m,
    \end{cases}.
  \]
  Direct simplification gives
  \[
  \begin{aligned}
  F(\boldsymbol\lambda_{m,4m-2})
    -\frac{4m^2-2m+8}{(4m^2-2m+2)^2}
    &=\frac{P_-(m)}
      {2(2m^2-m+1)^2(14m^3+63m^2-51m+25)^2},\\
  F(\boldsymbol\lambda_{m,4m})-\frac1{4(m^2-1)}
    &=\frac{P_+(m)}
      {4(m-1)(m+1)(14m^3+70m^2-23m+25)^2},
  \end{aligned}
  \]
  where
  \[
  \begin{aligned}
  P_-(m)&=(m-10)\bigl[m^3(200m^2-36m-800)
    +2148m^2+16258m+165231\bigr]+1653160,\\
  P_+(m)&=(m-10)\bigl[m(100m^2+180m-81)+1720\bigr]+9875.
  \end{aligned}
  \]
  Both bracketed expressions are positive for $m\ge10$, so
  $P_-(m),P_+(m)>0$.  Hence both spectra have purity strictly above the
  corresponding inner-polytope maximum.
\end{proof}

We next turn to the CAS maximum, which the central spectrum attains in every
bipartite dimension.

\section{CAS purity and absolute separability}
\label{sec:CAS-purity}

Throughout this section, $2\le m\le n$.
By the spectral characterization \cite[Theorem~1]{KBNKB}, every CAS state is
full rank, and for a full-rank state the CAS condition is equivalent to
\begin{equation}
  \label{eq:CAS-ratio}
  \frac{\lambda_{\max}}{\lambda_{\min}}
  \le \frac{m+1}{m-1}.
\end{equation}
For ordered spectra, \eqref{eq:CAS-ratio} is the linear inequality
$(m-1)\lambda_1\le(m+1)\lambda_d$.  The central spectrum has ratio
$(m+1)/(m-1)$ and is therefore CAS\@.

\begin{theorem}
  \label{thm:CAS-purity}
  For every $n\ge m\ge2$, the maximum purity of CAS states on
  $\BC^m\otimes\BC^n$ is $P^{\mathrm{cen}}_{m,n}$, and the
  unique maximizing ordered spectrum is
  $\boldsymbol{\lambda}^{\mathrm{cen}}_{m,n}$.  The maximizing CAS states are
  precisely the unitary conjugates of a diagonal state with this spectrum.
\end{theorem}

\begin{proof}
  Put $\kappa=(m+1)/(m-1)$.  By \eqref{eq:CAS-ratio}, the ordered CAS
  spectra form the compact polytope
  \[
    \{\boldsymbol\lambda\in\CO_d:\lambda_1\le\kappa\lambda_d\}.
  \]
  Normalization and $\lambda_1\le\kappa\lambda_d$ give
  $1\le d\kappa\lambda_d$, so $\lambda_d\ge1/(d\kappa)>0$.  Among the
  simplex vertices, only $\boldsymbol\nu_d$ satisfies this inequality.
  Hence every new vertex created by the halfspace cut lies on an edge
  $[\boldsymbol\nu_i,\boldsymbol\nu_d]$, $1\le i<d$, and satisfies
  $\lambda_1=\kappa\lambda_d$.

  These intersections are exactly the spectra
  $\boldsymbol\lambda_{i,d,\kappa}$ of Lemma~\ref{lem:fixed-ratio-purity}.
  Their purities are
  $\Phi_{d,\kappa}(i)=P(i)>1/d=F(\boldsymbol\nu_d)$.
  Lemma~\ref{lem:central-integer-maximum} gives the unique maximizing
  index $i_\star$ and value $P^{\mathrm{cen}}_{m,n}$.  Strict convexity
  excludes nonvertices, and unitary invariance gives all maximizing
  states.
\end{proof}

Recall $N=(m^2-1)n$.  Kondra et al.\ proved the universal CAS upper bound
$m/N$ \cite[Supplemental Material, Lemma~4]{KBNKB}.
By Lemma~\ref{lem:central-integer-maximum} and
Theorem~\ref{thm:CAS-purity}, the bound $m/N$ is attained exactly when
$(m-1)n$ is even.  When $(m-1)n$ is odd, the difference between this upper
bound and the exact maximum CAS purity is
\[
  \frac mN-P^{\mathrm{cen}}_{m,n}
  =\frac mN-\frac{m(N+2)}{(N+1)^2}
  =\frac{m}{N(N+1)^2}.
\]

For $m=2$ and $n\ge3$, the CAS maximum agrees with the exact qubit--qudit
APPT purity formulas of Song and Chen
\cite[Corollary~11(iv)]{SC}; our proof gives the CAS result directly for all
$n\ge2$.

For $m=3$, the CAS maximum is $3/(8n)$.  It agrees with the common maximum
purity of absolutely separable and APPT states for $n\ge9$, and is strictly
smaller for $3\le n\le8$ \cite[Theorem~A]{TrQ}.

\subsection{Consequences for absolute separability}

\begin{corollary}
  \label{cor:as-purity}
  Under the hypotheses of Theorem~\ref{thm:main}, APPT, AS, and CAS have
  a common maximum purity and common maximizing states.
\end{corollary}

\begin{proof}
  By Theorem~\ref{thm:CAS-purity}, the central spectrum
  $\boldsymbol{\lambda}^{\mathrm{cen}}_{m,n}$ is CAS\@.  Hence it belongs to
  AS and APPT\@.  The inclusions
  \[
    \mathrm{CAS}\subseteq\mathrm{AS}\subseteq\mathrm{APPT}
  \]
  and Theorem~\ref{thm:main} give the common maximum.  Unitary invariance
  and uniqueness of the maximizing ordered APPT spectrum then identify the
  common maximizing states.
\end{proof}

\section{Absolute separability of the inner polytope}
\label{sec:inner-polytope-asep}

Throughout this section, $2\le m\le n$.  For ordered eigenvalues, put
\[
 \begin{aligned}
 C_m(\boldsymbol{\lambda})
   &:=2\lambda_d+\sum_{j=1}^{m-1}\lambda_{d-j}
                   -\sum_{j=1}^{m-1}\lambda_j,\\
 \CP_{m,n}
   &:=\{\boldsymbol{\lambda}\in\CO_d:C_m(\boldsymbol{\lambda})\ge0\}.
 \end{aligned}
\]
Its permutation-invariant extension
\cite[Definition~6.1 and Theorem~6.2]{AKW} is
\[
 \widehat{\CP}_{m,n}
 :=\{\boldsymbol\lambda\in\Delta_{d-1}:
                C_m(\boldsymbol\lambda^\downarrow)\ge0\},
\]
where $\boldsymbol\lambda^\downarrow$ is the decreasing rearrangement.
For ordered spectra, the condition $C_m\ge0$ was known to imply APPT
\cite[Theorem~7.2]{J+}.  Theorem~\ref{thm:inner-polytope-asep}
strengthens this to absolute separability.  We prove it by checking
the vertices of the ordered polytope $\CP_{m,n}$.

\subsection{Proof of Theorem C}

\begin{proof}
The ordered set $\Lambda^{\mathrm{AS}}_{m,n}\cap\CO_d$ is convex:
convex combinations of ordered diagonal absolutely separable states remain
ordered, diagonal, and absolutely separable.  It therefore suffices to check
the vertices of $\CP_{m,n}$.

\smallskip \noindent\emph{Vertices of $\CP_{m,n}$.}
The simplex values are $C_m(\boldsymbol\nu_d)=2/d$,
$C_m(\boldsymbol\nu_{d-1})=0$, and
$C_m(\boldsymbol\nu_i)=-\chi_i/i<0$ for $1\le i\le d-2$, where
$\chi_i=\min\{i,m-1,d-1-i\}$.  Since only $\boldsymbol\nu_d$ has positive $C_m$-value,
Lemma~\ref{lem:active-face-dimension} shows that the vertices of
$\CP_{m,n}$ are $\boldsymbol\nu_d$, $\boldsymbol\nu_{d-1}$, and one
intersection on each edge $[\boldsymbol\nu_i,\boldsymbol\nu_d]$ with
$1\le i\le d-2$.
Writing this intersection as $\boldsymbol w_i=(a_i^{[i]},b_i^{[d-i]})$,
the equality $C_m(\boldsymbol w_i)=0$ and normalization give
$\chi_i(a_i-b_i)=2b_i$ and $i(a_i-b_i)+db_i=1$.  Hence
\begin{equation}
  \label{eq:inner-polytope-vertices}
 (a_i,b_i)=\frac{(\chi_i+2,\chi_i)}{d\chi_i+2i},
 \qquad 1\le i\le d-2.
\end{equation}
This recovers the vertices in
\cite[Eq.~(40) and the proof of Theorem~6.3]{AKW}.

\smallskip \noindent\emph{Absolute separability of the vertices.}
The vertices $\boldsymbol\nu_d$ and $\boldsymbol\nu_{d-1}$ have purity
at most $1/(d-1)$.  Since purity is unitary invariant, the
Gurvits--Barnum ball criterion \cite{GB} implies that they are absolutely separable.

For $1\le i\le m-1$, one has $\chi_i=i$, so
\eqref{eq:inner-polytope-vertices} gives $b_i=1/(d+2)$.  The
reduction-map criterion \cite[Theorem~3]{AVMRS} therefore proves absolute
separability for these $i$.

For $m\le i\le d-m$, one has $\chi_i=m-1$.  Hence the full-rank vertex
satisfies $a_i/b_i=(m+1)/(m-1)$ and is CAS by \eqref{eq:CAS-ratio}.

Finally, if $d-m+1\le i\le d-2$, put
$\ell=\chi_i=d-i-1\in[1,m-2]$.  Substitution in
\eqref{eq:inner-polytope-vertices} gives
\[
 F(\boldsymbol w_i)
 =\frac1{d-1}-
 \frac{d\ell^2}{(d-1)(\ell(d-2)+2d-2)^2}<\frac1{d-1}.
\]
These vertices are absolutely separable by \cite{GB}.  This checks every
ordered vertex, so convexity gives
$\CP_{m,n}\subseteq
\Lambda^{\mathrm{AS}}_{m,n}\cap\CO_d$, and permutation invariance proves
$\widehat{\CP}_{m,n}\subseteq\Lambda^{\mathrm{AS}}_{m,n}$.
\end{proof}

For $m=2$, Theorem~\ref{thm:inner-polytope-asep} already follows from
$\Lambda^{\mathrm{AS}}_{2,n}=\Lambda^{\mathrm{APPT}}_{2,n}$ \cite{Jo}; its
new content is for $m\ge3$.

\subsection{Comparison with earlier linear criteria}

Here \emph{linear} means linear in the decreasingly ordered eigenvalues.
The reduction-map regions of \cite[Theorem~3]{AVMRS} are
\[
 \mathcal B_+
 :=\left\{\boldsymbol{\lambda}\in\Delta_{d-1}:
                    \lambda_{\min}\ge\frac1{d+2}\right\},
 \qquad
 \mathcal B_-
 :=\left\{\boldsymbol{\lambda}\in\Delta_{d-1}:
                    \lambda_{\max}\le\frac1{d-1}\right\}.
\]
For decreasingly ordered spectra, their convex hull is described by the single
inequality \cite[Lemma~1 and Theorem~4]{AVMRS}
\begin{equation}
  \label{eq:earlier-linear-criterion}
 3\sum_{i=d-q+1}^d\lambda_i
 +(d+2-3q)\lambda_{d-q}\ge1,
 \qquad q:=\left\lceil\frac{d-1}{3}\right\rceil.
\end{equation}
The other comparison region is $\Lambda^{\mathrm{CAS}}_{m,n}$, defined by
the ratio inequality \eqref{eq:CAS-ratio}.  Let
\[
 \mathcal K_{m,n}
 :=\operatorname{conv}
   (\mathcal B_+\cup\mathcal B_-\cup\Lambda^{\mathrm{CAS}}_{m,n}).
\]

\begin{proposition}
\label{prop:inner-polytope-comparison}
For every $2\le m\le n$,
\begin{equation}\label{eq:linear-region-inclusions}
 \mathcal K_{m,n}\subseteq\widehat{\CP}_{m,n}
 \subseteq\Lambda^{\mathrm{AS}}_{m,n}.
\end{equation}
Both $\operatorname{conv}(\mathcal B_+\cup\mathcal B_-)$ and
$\Lambda^{\mathrm{CAS}}_{m,n}$ are proper subsets of $\widehat{\CP}_{m,n}$.
The first inclusion in \eqref{eq:linear-region-inclusions} is an
equality when $m=2$ and is strict when $m\ge3$.
\end{proposition}

\begin{proof}
By permutation invariance, we work with decreasingly ordered spectra and write
$C_m(\boldsymbol{\lambda})=\sum_i c_i\lambda_i$.  The coefficients are
$-1$ on the first $m-1$ positions, $0$ in the middle, $1$ on the next
$m-1$ positions, and $2$ at the last position; in particular, they are
nondecreasing and sum to $2$.  The rearrangement inequality gives
$C_m(\boldsymbol{\lambda}^\downarrow)
=\min_\pi\sum_i c_i\lambda_{\pi(i)}$.  Thus $\widehat{\CP}_{m,n}$ is an
intersection of halfspaces and is convex.  Moreover,
\[
 \begin{aligned}
 C_m(\boldsymbol{\lambda})
 &=\sum_i(c_i+1)\lambda_i-1
   \ge(d+2)\lambda_d-1,\\
 C_m(\boldsymbol{\lambda})
 &=2-\sum_i(2-c_i)\lambda_i
   \ge2-2(d-1)\lambda_1.
 \end{aligned}
\]
Thus $\mathcal B_+,\mathcal B_-\subseteq\widehat{\CP}_{m,n}$.  The CAS
ratio inequality gives
$C_m(\boldsymbol{\lambda})\ge(m+1)\lambda_d-(m-1)\lambda_1\ge0$, so
$\Lambda^{\mathrm{CAS}}_{m,n}\subseteq\widehat{\CP}_{m,n}$.
Convexity and Theorem~\ref{thm:inner-polytope-asep} now prove
\eqref{eq:linear-region-inclusions}.

When $m=2$, one has $\chi_i=1$ in
\eqref{eq:inner-polytope-vertices}, so every edge vertex
$\boldsymbol w_i$ is CAS\@.  The remaining full-rank vertex
$\boldsymbol\nu_d$ is also CAS, while
$\boldsymbol\nu_{d-1}\in\mathcal B_-$.  Thus every vertex of
$\CP_{2,n}$ lies in $\mathcal K_{2,n}$, and permutation invariance gives
$\widehat{\CP}_{2,n}\subseteq\mathcal K_{2,n}$.  Hence equality holds
when $m=2$.

The spectrum $\boldsymbol\nu_{d-1}=((1/(d-1))^{[d-1]},0)$ belongs to
$\mathcal B_-$ and hence to $\widehat{\CP}_{m,n}$, but is not CAS because
it is rank deficient.  Thus
$\Lambda^{\mathrm{CAS}}_{m,n}$ is a proper subset.  To show that
$\operatorname{conv}(\mathcal B_+\cup\mathcal B_-)$ is also proper, take
\[
 \boldsymbol\eta
 :=\frac{\bigl((m+1)^{[m]},(m-1)^{[d-m]}\bigr)}{(m-1)d+2m}.
\]
Its largest-to-smallest eigenvalue ratio is $(m+1)/(m-1)$, so it is CAS\@.
Since $d\ge m^2$ implies $q+1\le d-m$, every eigenvalue appearing
in the left-hand side of \eqref{eq:earlier-linear-criterion} equals the smaller
eigenvalue of $\boldsymbol\eta$, and
the left-hand side is
\[
 \frac{(d+2)(m-1)}{(m-1)d+2m}
 =1-\frac2{(m-1)d+2m}<1.
\]
Hence $\boldsymbol\eta$ lies outside
$\operatorname{conv}(\mathcal B_+\cup\mathcal B_-)$.

Finally, let $m\ge3$ and consider
\begin{equation}
 \boldsymbol{\lambda}^{\,\dagger}
 =\frac{\bigl(3^{[d-2]},1,1\bigr)}{3d-4}.
\end{equation}
Since $C_m(\boldsymbol{\lambda}^{\,\dagger})=0$, we have
$\boldsymbol{\lambda}^{\,\dagger}\in\widehat{\CP}_{m,n}$.  To show that it
lies outside $\mathcal K_{m,n}$, put
$\tau(\boldsymbol\lambda):=\min_{i<j}(\lambda_i+\lambda_j)$.
Since $\tau$ is the minimum of the finitely many linear functions
$\lambda_i+\lambda_j$, it is concave and permutation invariant.  For an
ordered spectrum, $\tau=\lambda_{d-1}+\lambda_d$.

Each of the three sets defining $\mathcal K_{m,n}$ satisfies
$\tau\ge1/(d-1)$.  On $\mathcal B_+$ this follows from
$\tau\ge2/(d+2)$.  On $\mathcal B_-$,
$\tau=1-\sum_{i=1}^{d-2}\lambda_i\ge1/(d-1)$.  For a CAS spectrum,
the ratio criterion gives
\[
  \lambda_1\le\frac{m+1}{m-1}\lambda_d
  \le2\lambda_d\le\tau,
\]
where the second inequality uses $m\ge3$.  Hence the first $d-2$
eigenvalues are at most $\tau$, and normalization gives
$1\le(d-1)\tau$.  Because $\tau$ is concave, the inequality
$\tau\ge1/(d-1)$ is preserved under convex combinations.  Hence it holds
throughout $\mathcal K_{m,n}$.

But $\tau(\boldsymbol{\lambda}^{\,\dagger})=2/(3d-4)<1/(d-1)$.  Therefore
$\boldsymbol{\lambda}^{\,\dagger}\notin\mathcal K_{m,n}$, and
$\mathcal K_{m,n}\subsetneq\widehat{\CP}_{m,n}$ for $m\ge3$.
\end{proof}

\section{Relative spectral volumes}
\label{sec:volume}

By Theorem~\ref{thm:inner-polytope-asep}, the inner polytope satisfies
$\widehat{\CP}_{m,n}\subseteq\Lambda^{\mathrm{AS}}_{m,n}$, while the
necessary APPT inequality $G_m\ge0$ gives
$\Lambda^{\mathrm{APPT}}_{m,n}\subseteq\widehat{\CU}_{m,n}$.
Thus the known exact volume of $\widehat{\CP}_{m,n}$
\cite[Proposition~6.14]{AKW} gives a lower bound for the AS and APPT
spectral volumes.  In this section we additionally compute the exact CAS
volume and the outer-polytope volume, together with their fixed-$m$
asymptotics.
Throughout this section, $2\le m\le n$, and we put $D=d-m$.
Recall that
\[
 d=mn,\qquad
 r_m=\frac{m(m-1)}2,\qquad s_m=\frac{m(m+1)}2.
\]
For $\boldsymbol{\lambda}\in\Delta_{d-1}$, write
$\lambda_1^\downarrow\ge\cdots\ge\lambda_d^\downarrow$ for its
decreasing rearrangement and define the permutation-invariant outer
polytope
\begin{equation}
  \label{eq:full-one-cut-outer}
 \widehat{\CU}_{m,n}
 :=\left\{\boldsymbol{\lambda}\in\Delta_{d-1}:
 \sum_{i=1}^{r_m}\lambda_i^\downarrow
 \le \sum_{i=0}^{s_m-1}\lambda_{d-i}^\downarrow\right\}.
\end{equation}
The sum on the left is the maximum over all $r_m$-coordinate sums,
while the sum on the right is the minimum over all $s_m$-coordinate sums.
Their head and tail index sets are disjoint because
$r_m+s_m=m^2\le d$.  Hence \eqref{eq:full-one-cut-outer} is equivalent to
$\sum_{i\in A}\lambda_i\le\sum_{j\in B}\lambda_j$ for every disjoint
pair with $|A|=r_m$ and $|B|=s_m$, so it is a polytope defined by finitely
many linear inequalities.  The $k=m$ case of Lemma~\ref{lem:head-tail}
gives
\[
 \Lambda^{\mathrm{APPT}}_{m,n}\subseteq\widehat{\CU}_{m,n}.
\]
Its intersection with the ordered simplex is
$\{\boldsymbol\lambda\in\CO_d:G_m(\boldsymbol\lambda)\ge0\}$.

For a measurable subset $K\subseteq\Delta_{d-1}$, we put
\[
 \volrel(K)
 :=\frac{\operatorname{vol}_{d-1}(K)}
 {\operatorname{vol}_{d-1}(\Delta_{d-1})}.
\]
Here $\operatorname{vol}_{d-1}$ is Euclidean measure on the affine
hyperplane $\sum_i\lambda_i=1$.  For a permutation-invariant set $K$,
symmetry gives
\[
 \volrel(K)
 =\frac{\operatorname{vol}_{d-1}(K^\downarrow)}
 {\operatorname{vol}_{d-1}(\CO_d)},
 \qquad K^\downarrow:=K\cap\CO_d.
\]

\subsection{Exact CAS and outer volumes}

\begin{proposition}
For every $n\ge m\ge2$,
\begin{align*}
 V^{\mathrm{CAS}}_{m,n}
 :=\volrel(\Lambda^{\mathrm{CAS}}_{m,n})
 &=\frac{(m-1)d}{2}B\!\left(d,\frac{(m-1)d}{2}\right)
 \\
 &=\frac{\Gamma(d)\Gamma\!\left(\frac{m-1}{2}d+1\right)}
 {\Gamma\!\left(\frac{m+1}{2}d\right)},
\end{align*}
where $B$ and $\Gamma$ are Euler's beta and gamma functions.
\end{proposition}

\begin{proof}
The proof of Theorem~\ref{thm:CAS-purity} identifies the ordered CAS set
as the simplex with vertices $\boldsymbol\nu_d$ and
\[
 \boldsymbol\nu_d+\frac{2i}{(m-1)d+2i}
       (\boldsymbol\nu_i-\boldsymbol\nu_d),\qquad 1\le i<d.
\]
Relative to $\CO_d=[\boldsymbol\nu_1,\ldots,\boldsymbol\nu_d]$,
the $i$th edge vector from $\boldsymbol\nu_d$ is scaled by
$2i/((m-1)d+2i)$.  Since scaling the $i$th edge vector of a simplex
by a factor $\alpha_i$ multiplies its volume by
$\prod_i\alpha_i$, the relative volume of the ordered CAS simplex in
$\CO_d$ is
\[
 \prod_{i=1}^{d-1}\frac{2i}{(m-1)d+2i}.
\]
Permutation symmetry gives the same relative volume in $\Delta_{d-1}$.  Hence
\[
 V^{\mathrm{CAS}}_{m,n}
 =\prod_{i=1}^{d-1}\frac{2i}{(m-1)d+2i}
 =\frac{\Gamma(d)\Gamma\!\left(\frac{m-1}{2}d+1\right)}
        {\Gamma\!\left(\frac{m+1}{2}d\right)}.
\]
The identity $B(x,y)=\Gamma(x)\Gamma(y)/\Gamma(x+y)$ gives the beta-function
form.
\end{proof}

The outer-volume formula uses only the $k=m$ inequality in
Lemma~\ref{lem:head-tail}.

\begin{proposition}
\label{prop:outer-relative-volume}
For $n\ge m\ge2$, set
\[
 \zeta_\ell=\frac{\ell}{D+\ell}\qquad(1\le \ell\le m).
\]
Then the relative volume of the outer polytope
\eqref{eq:full-one-cut-outer} is
\begin{equation}
  \label{eq:outer-volume-product}
\begin{split}
 V^{\mathrm{out}}_{m,n}
 &:=\volrel(\widehat{\CU}_{m,n})\\
 &=\sum_{\ell=1}^{m}
 \frac{\zeta_\ell^{d-1}}
 {\zeta_\ell(\zeta_\ell+1)^{r_m}
  \displaystyle\prod_{i=r_m+1}^{d-s_m}\left(\zeta_\ell+\frac{r_m}{i}\right)
  \displaystyle\prod_{j=1}^{r_m-1}\left(\zeta_\ell+\frac{j}{D-j}\right)
  \displaystyle\prod_{\substack{1\le q\le m\\q\ne \ell}}
       (\zeta_\ell-\zeta_q)}.
\end{split}
\end{equation}
Here an empty product is $1$.
\end{proposition}

\begin{proof}
Write
$\boldsymbol\lambda=\sum_{i=1}^d\theta_i\boldsymbol\nu_i$ with
$\boldsymbol\theta\in\Delta_{d-1}$.  This affine bijection multiplies all
$(d-1)$-dimensional volumes by the same constant factor and hence preserves
relative volumes.  The inequality $G_m(\boldsymbol\lambda)\ge0$ becomes
$\sum_iG_m(\boldsymbol\nu_i)\theta_i\ge0$.
By \eqref{eq:Gk-simplex-values}, with $D=d-m$, these values are
\[
 \begin{aligned}
 G_m(\boldsymbol\nu_i)&=-1
     &&(1\le i\le r_m),\\
 G_m(\boldsymbol\nu_i)&=-\frac{r_m}{i}
     &&(r_m<i\le d-s_m),\\
 G_m(\boldsymbol\nu_{D-j})&=-\frac{j}{D-j}
     &&(1\le j<r_m),\\
 G_m(\boldsymbol\nu_D)&=0,\qquad
 G_m(\boldsymbol\nu_{D+q})=\zeta_q
     &&(1\le q\le m).
 \end{aligned}
\]
Applying the simplex-section formula of \cite{La}, recalled in
\cite[Theorem~6.13]{AKW}, to the halfspace
$\sum_iG_m(\boldsymbol\nu_i)\theta_i\ge0$ in $\Delta_{d-1}$ gives the
relative volume
\[
 \sum_{\ell=1}^m
 \frac{\zeta_\ell^{d-1}}
 {\prod_{i\ne D+\ell}
   \bigl(\zeta_\ell-G_m(\boldsymbol\nu_i)\bigr)}.
\]
Some of the values $G_m(\boldsymbol\nu_i)$ coincide, but the perturbation
argument of \cite[Section~2.1]{La} extends the simplex-section formula to
this situation.  For fixed $\ell$, the value
$G_m(\boldsymbol\nu_D)=0$ contributes $\zeta_\ell$, the values equal to
$-1$ contribute $(\zeta_\ell+1)^{r_m}$, the two remaining negative
families give the corresponding products in \eqref{eq:outer-volume-product},
and the positive values $G_m(\boldsymbol\nu_{D+q})=\zeta_q$ with
$q\ne\ell$ give $\prod_{q\ne\ell}(\zeta_\ell-\zeta_q)$.
Substituting these factors into the denominator yields
\eqref{eq:outer-volume-product}.
By permutation symmetry, this relative volume equals
$\volrel(\widehat{\CU}_{m,n})$.
\end{proof}

When $m=2$, $G_m=G_2$, so $\widehat{\CU}_{2,n}$ is exactly the
symmetrized outer polytope studied in \cite{TrS}.  Accordingly,
Proposition~\ref{prop:outer-relative-volume} reduces to the beta-function
formula for the qubit--qudit outer relative volume in \cite{TrS}.

\subsection{Leading volume asymptotics}

Define
\[
 \beta_m^{\mathrm{out}}
 :=\frac{r_m^{\,r_m}s_m^{\,s_m-1}}{r_m!(s_m-1)!}.
\]

\begin{lemma}
\label{lem:volume-asymptotics}
For fixed $m\ge2$, as $n\to\infty$ with $d=mn$,
\begin{align}
 V^{\mathrm{CAS}}_{m,n}
 &=\sqrt{\frac{\pi}{2}(m^2-1)d}\,
   e^{-\kappa_m n}\bigl(1+O_m(d^{-1})\bigr),
 \label{eq:cas-leading-asymptotic}\\
 \frac{V^{\mathrm{out}}_{m,n}}{V^{\mathrm{CAS}}_{m,n}}
 &=\beta_m^{\mathrm{out}}\bigl(1+O_m(d^{-1})\bigr).
 \label{eq:outer-cas-ratio}
\end{align}
\end{lemma}

\begin{proof}
Put $a=(m-1)/2$.  Since
$m[a\log a-(a+1)\log(a+1)]=-\kappa_m$, Stirling's formula gives
\[
 \frac{\Gamma(d)\Gamma(ad+1)}{\Gamma((a+1)d)}
 =\sqrt{2\pi a(a+1)d}\,e^{-\kappa_m n}
   \bigl(1+O_m(d^{-1})\bigr),
\]
which proves \eqref{eq:cas-leading-asymptotic}.

Let $T_\ell$ denote the $\ell$th summand in
\eqref{eq:outer-volume-product}.  We first consider $T_m$, which gives the
leading contribution.

For $\ell=m$, we use $\zeta_m=m/d$ and
$V^{\mathrm{CAS}}_{m,n}=\prod_{i=1}^{d-1}mi/(r_m d+mi)$.
Cancelling the factors with $r_m<i\le d-s_m$ gives
\begin{equation}
  \label{eq:dominant-volume-ratio}
 \frac{T_m}{V^{\mathrm{CAS}}_{m,n}}
 =\frac{\displaystyle\prod_{i=1}^{r_m}(r_m d+mi)}
        {r_m!(d+m)^{r_m}}
  \frac{\displaystyle\prod_{j=1}^{s_m-1}(s_m d-mj)}
        {(s_m-1)!(d-m)^{s_m-1}}.
\end{equation}
The two factors tend to $r_m^{r_m}/r_m!$ and
$s_m^{s_m-1}/(s_m-1)!$, respectively, with relative error
$O_m(d^{-1})$.  Hence
$T_m/V^{\mathrm{CAS}}_{m,n}
=\beta_m^{\mathrm{out}}(1+O_m(d^{-1}))$.

For $\ell<m$ and each index $i$ with
$G_m(\boldsymbol\nu_i)<0$, set
\[
 R_i:=\frac{\zeta_\ell
   (\zeta_m-G_m(\boldsymbol\nu_i))}
 {\zeta_m(\zeta_\ell-G_m(\boldsymbol\nu_i))}.
\]
Using $\zeta_q=q/(D+q)$ and
$\zeta_a-\zeta_b=D(a-b)/[(D+a)(D+b)]$ to simplify the factors arising
from the positive values $G_m(\boldsymbol\nu_{D+q})=\zeta_q$ gives
\[
 \left|\frac{T_\ell}{T_m}\right|
 =\binom{m-1}{\ell-1}
   \left(\frac{\ell}{m}\right)^{m-1}\frac d{D+\ell}
   \prod_{\substack{1\le i\le d\\G_m(\boldsymbol\nu_i)<0}}R_i.
\]
Since $(\ell/m)^{m-1}\le1$ and
$d/(D+\ell)\le d/(d-m+1)\le m/(m-1)$, the prefactor is at most
$\frac m{m-1}\binom{m-1}{\ell-1}$.  Moreover, $R_i\le1$ whenever
$G_m(\boldsymbol\nu_i)<0$, because $0<\zeta_\ell<\zeta_m$.
For the $d-m^2$ indices $r_m<i\le d-s_m$, where
$G_m(\boldsymbol\nu_i)=-r_m/i$, the sharper bound is
\[
 R_i
 =\frac{\ell(mi+r_md)}{m(\ell i+r_m(D+\ell))}
 \le\frac{\ell s_m}{m(r_m+\ell)}
 \le\frac{m+1}{m+2}<1.
\]
The first inequality uses $i\le d-s_m$; the second uses $\ell\le m-1$.
Summing over $\ell<m$ therefore gives, for every $n\ge m$,
\[
 \left|\frac{V^{\mathrm{out}}_{m,n}}{T_m}-1\right|
 \le\frac m{m-1}(2^{m-1}-1)\left(\frac{m+1}{m+2}\right)^{d-m^2}.
\]
Together with \eqref{eq:dominant-volume-ratio}, this proves
\eqref{eq:outer-cas-ratio}.
\end{proof}

\begin{proof}[Proof of Theorem~\ref{thm:volume-main}]
The spectral inclusions and Lemma~\ref{lem:head-tail} give
\begin{equation}
  \label{eq:volume-bounds}
 V^{\mathrm{CAS}}_{m,n}
 \le\volrel(\Lambda^{\mathrm{AS}}_{m,n})
 \le\volrel(\Lambda^{\mathrm{APPT}}_{m,n})
 \le V^{\mathrm{out}}_{m,n}.
\end{equation}
Lemma~\ref{lem:volume-asymptotics} shows that the two bounding volumes
have order $\sqrt n\,e^{-\kappa_m n}$ up to positive constants depending
only on $m$.  Hence every volume in \eqref{eq:volume-bounds} has the same
order.  This proves the $\Theta_m$-statement, and taking logarithms gives
the equivalent logarithmic form.  Finally, with $a=(m-1)/2>0$,
$\kappa_m=m[\log(a+1)+a\log(1+1/a)]>0$.
\end{proof}
\section{Conclusion}

We have determined the exact APPT purity maximum in the large-$n$ ranges of
Theorem~\ref{thm:main}, constructed explicit APPT boundary spectra
that disprove both cases of Conjecture~\ref{conj:AKW} for every $m\ge10$,
and determined the exact CAS purity maximum in every bipartite dimension.

The inner APPT polytope consists entirely of absolutely separable spectra.
Its linear criterion strictly improves both the reduction-map criterion and
the CAS ratio criterion; for $m\ge3$, it also strictly contains the convex
hull of the two reduction-map regions and the CAS region.  The exact CAS and
outer-polytope volume formulas show that the CAS, AS, and APPT volumes have the
same asymptotic order $\sqrt n\,e^{-\kappa_m n}$ for each fixed $m$.

The common maximum purity of absolutely separable and APPT states for
$m=3$ is determined for every $n\ge3$ in \cite{TrQ}.  For $m\ge4$, the
exact APPT purity maximum below the dimension bounds in
\eqref{eq:main-dimension-bounds} remains open.  The counterexample spectra in
Theorem~\ref{thm:counterexample-families} provide lower bounds on the maximum
but need not be global maximizers.  Optimizing over the boundary family in
Proposition~\ref{prop:boundary-spectra} may help determine
these remaining maxima.

For each fixed $m\ge4$, a related problem is to determine the least integer
$n_0$ for which the central spectrum is the unique APPT purity maximizer
for every $n\ge n_0$.  Theorem~\ref{thm:main} gives an upper bound on $n_0$.
That bound cannot be lowered using $\CR_{m,n}$: its purity maximum exceeds
the central value when $n$ is one below the bound
(Subsection~\ref{sec:outer-sharpness}).  Improving this dimension bound
therefore requires APPT information beyond the two inequalities defining
$\CR_{m,n}$.

The exact AS and APPT volumes remain unknown.  Determining them, or their
asymptotics relative to CAS, would clarify the relative sizes of the three
sets.  By Theorem~\ref{thm:inner-polytope-asep}, the known exact volume of
$\widehat{\CP}_{m,n}$ gives a lower bound for both volumes, while the
outer-polytope volume gives an upper bound.  The exact CAS volume obtained
here provides a further explicit comparison with the inner-polytope lower
bound and the outer-polytope upper bound.

\section*{Declarations}

\smallskip \noindent\textbf{Funding.}
The author received no funding for this work.

\medskip
\noindent\textbf{Competing interests.}
The author declares no competing interests.

\medskip
\noindent\textbf{Use of AI tools.}
ChatGPT 5.6 Sol (OpenAI) was used to perform symbolic computations and
simplify some mathematical arguments, and ChatGPT 6 was used to improve
the exposition.  Computations were also cross-checked with Mathematica.
The author verified all mathematical content and takes full responsibility
for the manuscript.

\medskip
\noindent\textbf{Data availability.}
No datasets were generated or analyzed.

\end{document}